\documentclass[aps,prl,twocolumn,a4,showkeys,10pt,longbibliography,superscriptaddress]{revtex4-2}

\usepackage{graphicx}
\usepackage[english]{babel}
\usepackage{amsmath}
\usepackage{float}
\usepackage{hyperref}
\usepackage{caption}
\usepackage{subcaption}
\usepackage{tikz}
\usepackage{tikzsymbols}
\usepackage{physics}
\usepackage{breqn}
\usepackage{mathtools}
\usepackage{enumerate}
\usepackage{amssymb}
\usepackage{bbold}
\usepackage{ragged2e}
\usepackage[normalem]{ulem} 
\usepackage{makecell} 
\usepackage{multirow}

\makeatletter
\long\def\@makecaption#1#2{%
  \par
  \vskip\abovecaptionskip
  \sbox\@tempboxa{#1: #2}%
  \ifdim \wd\@tempboxa >\hsize
    \justifying #1: #2\par 
  \else
    \global \@minipagefalse
    \hb@xt@\hsize{\hfil\box\@tempboxa\hfil}%
  \fi
  \vskip\belowcaptionskip
}
\makeatother

\newcommand{\sev}[1]{\langle #1 \rangle}

\begin{document}

\title{Continuous feedback protocols for cooling and trapping a quantum harmonic oscillator}
\date{\today}

\author{Guilherme De Sousa} \email{gdesousa@umd.edu}
\affiliation{Department of Physics$,$ University of Maryland$,$ College Park$,$ MD 20742$,$ USA}

\author{Pharnam Bakhshinezhad}
\thanks{Formerly known as Faraj Bakhshinezhad.}
\affiliation{Atominstitut$,$ Technische Universität Wien$,$ Stadionallee 2$,$ 1020 Vienna$,$ Austria}
\affiliation{Physics Department and NanoLund$,$ Lund University$,$ Box 118$,$ 22100 Lund$,$ Sweden}

\author{Bj{\"o}rn Annby-Andersson}
\affiliation{Physics Department and NanoLund$,$ Lund University$,$ Box 118$,$ 22100 Lund$,$ Sweden}

\author{Peter Samuelsson}
\affiliation{Physics Department and NanoLund$,$ Lund University$,$ Box 118$,$ 22100 Lund$,$ Sweden}

\author{Patrick P. Potts}
\affiliation{Department of Physics$,$ University of Basel$,$ Klingelbergstrasse 82$,$ 4056 Basel$,$ Switzerland}

\author{Christopher Jarzynski}
\affiliation{Department of Physics$,$ University of Maryland$,$ College Park$,$ MD 20742$,$ USA}
\affiliation{Institute for Physical Science and Technology$,$ University of Maryland$,$ College Park$,$ MD 20742$,$ USA}

\begin{abstract}
Quantum technologies and experiments often require preparing systems in low-temperature states. Here, we investigate cooling schemes using feedback protocols modeled with a Quantum Fokker-Planck Master Equation (QFPME) recently derived by Annby-Andersson et.~al. ({\it Phys.~Rev.~Lett.} \textbf{129}, 050401, 2022).  This equation describes systems under continuous weak measurements, with feedback based on the outcome of these measurements.  We apply this formalism to study the cooling and trapping of a harmonic oscillator for several protocols based on position and/or momentum measurements. We find that the protocols can cool the oscillator down to, or close to, the ground state for suitable choices of parameters. 
Our analysis provides an analytically solvable case study of quantum measurement and feedback and illustrates the application of the QFPME to continuous quantum systems.
\end{abstract}

\maketitle
\section{Introduction}
Technologies and techniques to manipulate individual quantum systems have advanced rapidly in recent years. Applications such as quantum computing \cite{HAFFNER2008155,RevModPhys.82.2313,doi:10.1126/science.1231930} and sensing \cite{RevModPhys.89.035002,Pirandola2018} rely on fine measurements and precise control, along with the ability to prepare systems in desired states. There is particular interest in cooling systems to near-zero temperatures, that is, near the ground state, where quantum phenomena typically become enhanced. 

Standard techniques for cooling quantum systems include laser cooling \cite{doi:10.1126/science.239.4842.877,doi:10.1063/1.881239,doi:10.1126/science.253.5022.861,RevModPhys.70.721}, evaporative cooling \cite{PhysRevLett.74.5202,Davis1995,Ketterle1996,PhysRevA.78.011604} and cavity cooling \cite{Maunz2004,doi:10.1073/pnas.1309167110,PhysRevLett.118.183601}.
Alternatively, systems can be cooled through measurement and feedback \cite{PhysRevLett.90.043001,PhysRevLett.96.043003,PhysRevB.68.235328,PhysRevA.91.043812,Schafermeier2016,Habibi_2016,PhysRevLett.123.223602,PhysRevLett.122.070603,PhysRevLett.117.163601}, in a manner reminiscent of the Maxwell's demon thought experiment.
Fundamental costs and bounds associated with cooling quantum systems are subjects of active investigation \cite{taranto2024cooling,https://doi.org/10.48550/arxiv.2106.05151, PhysRevApplied.14.054005,Freitas2018}.

\begin{figure}[!h]
    \centering
    \includegraphics[width=0.8\linewidth]{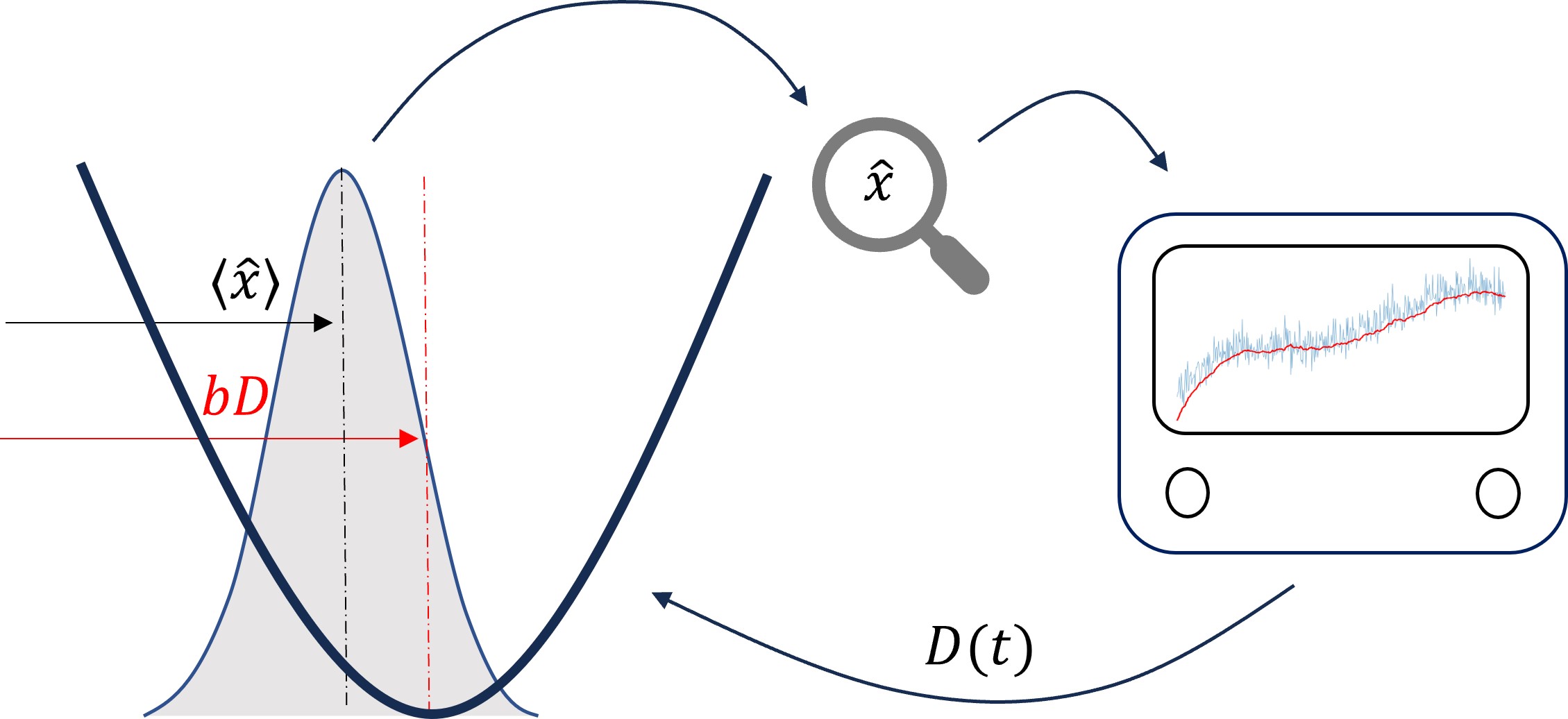}
    \caption{Illustration of Protocol X (defined in the next section). Cooling is achieved by measuring the position continuously and applying feedback by changing the location of the harmonic potential. The measurement outcome is a low-pass filtered version (red line) of the noisy weak measurement (light blue line).}
    \label{fig:prot1_diagram}
\end{figure}

Recently Annby-Andersson et.~al.~\cite{PhysRevLett.129.050401} introduced a theoretical framework for modeling continuous, Markovian measurement and feedback of quantum systems. In this framework, an external agent continuously monitors a system observable $\hat A$ using weak measurements \cite{PhysRevA.36.5543,DIOSI1988419,Dyrting_1996,PhysRevA.62.012105,doi:10.1126/science.1095374,doi:10.1080/00107510601101934,wiseman_milburn_2009,PhysRevLett.104.093601,doi:10.1126/science.1225258,ZHANG20171,Minev2019}. The noisy measurement outcome $D(t)$, supplied by a detector of finite bandwidth, enters as a parameter of the system's Hamiltonian (or Liouvillean for an open system).
In this manner, feedback is applied to the system, as seen in Fig.~\ref{fig:prot1_diagram}.
The central result of Ref.~\cite{PhysRevLett.129.050401} is the Quantum Fokker-Planck Master Equation (QFPME) describing, at the ensemble level, the coupled dynamics of the quantum system and detector outcome.

Here, we apply the framework of Ref.~\cite{PhysRevLett.129.050401} to study the cooling of a quantum harmonic oscillator under various feedback protocols. 
We imagine that an external agent continuously performs weak measurements of the system's position, $x$, or momentum, $p$, or both, and manipulates the location of the harmonic well based on the measurement outcomes.
The agent aims to bring the system to the harmonic oscillator ground state, centered at the origin, $x=0$.
We will see that the goals of cooling (to the ground state) and trapping (at the origin) compete with one another.
This system is simple enough to obtain analytical results, allowing us to compare the asymptotic energy under different protocols and to investigate quantitatively how the rate and degree of cooling depend on parameters such as the detector bandwidth, the oscillator's natural frequency, and the measurement strength.

At the same time, the cooling of a quantum harmonic oscillator is experimentally relevant and has been proposed and/or implemented in trapped ions, nano- and micro-mechanical resonators, levitated nanoparticles and optomechanical cavities~\cite{PhysRevLett.85.4458,PhysRevLett.92.075507,PhysRevA.83.043804,Teufel2011,Chan2011,PhysRevX.2.041014,PhysRevLett.117.163601,Zeng2017,PhysRevLett.90.043001,PhysRevLett.96.043003,PhysRevB.68.235328,PhysRevLett.123.223602}.
The results presented in this work differ from previous studies of continuous feedback in harmonic oscillators \cite{PhysRevA.76.013610,PhysRevE.105.044137,https://doi.org/10.48550/arxiv.2212.12292,doi:10.1098/rsta.2011.0531} due to the presence of finite bandwidth. This built-in feature of the QFPME \cite{PhysRevLett.129.050401} allows for feeding back non-linear functions of the measurement outcome.

Exploiting the Markovianity of the QFPME and the system's symmetries, we derive closed formulas for the asymptotic energy of a quantum harmonic oscillator under continuous feedback cooling. We show that by tuning the protocol, one can obtain either a cooling-focused protocol or a trapping-focused protocol. If the system is weakly coupled to a thermal bath, its final energy is an average of the asymptotic feedback energy and the thermal energy, weighted  by the cooling and thermal relaxation rates.

This paper is structured as follows. In Sec.~\ref{sec:model}, we introduce the harmonic oscillator model and cooling protocols, and we briefly summarize the QFPME of Ref.~\cite{PhysRevLett.129.050401}. Section~\ref{sec:results} describes the cooling and trapping results achieved using these protocols. Section~\ref{sec:trajectory} presents numerical simulations and compares them with analytical results. Finally, we compare protocols in Sec.~\ref{sec:discussion} and conclude in Sec.~\ref{sec:conclusion}.

\section{Model and QFPME}\label{sec:model}
We consider a system that is continuously measured in one or more observables. Based on the outcomes of these measurements, which are generally noisy, an external agent implements a time-dependent Hamiltonian with the goal of decreasing the system's internal energy. We will refer to this process as a {\it cooling protocol}.  As described above, a secondary aim is to trap the particle at or near the origin, in the long time limit.

\subsection{Classical model}\label{sec:classical_model}
Although this paper focuses on quantum feedback control, it is useful to gain intuition by first analyzing a classical cooling protocol. Consider a classical harmonic oscillator described by the Hamiltonian 
\begin{equation}
    \label{eq:Hamiltonian_classical_osc}
    H= \frac{p^2}{2m} + \frac{m\omega^2}{2}(x - bD_x)^2.
\end{equation}
Here, $D_x$ represents a recent measurement of the particle's position, and $b>0$ is a fixed parameter that tunes how the external agent manipulates the harmonic well's location in response to this measurement.
In the following, we describe the behavior of this system under both discrete and continuous cooling protocols. Details of the analyses of these protocols are found in Appendix~\ref{sec:A_classical_model}.

We first consider a cooling protocol that proceeds in discrete time steps $\Delta t$, where $\Delta t$ is not an integer multiple of $\pi/\omega$.  At each step, the agent measures the particle's current position (i.e.\ $D_x=x$) and instantaneously moves the trap's location to $bD_x$.
Between measurements the particle evolves under the  Hamiltonian $H$, with $D_x$ fixed.
If $b=1$, the particle loses all of its potential energy at each step, as the minimum of the harmonic potential is moved to the particle's current location.
At long times, the particle comes to rest at the minimum of the harmonic well, at a location $D_x^\infty\ne 0$ that depends on the initial position and momentum $(x_0,p_0)$ (see Eq.~\eqref{eq:asymp-discrete}).  Thus the particle is cooled but is not trapped at the origin.
By contrast, if $b<1$ then at each time step the trap is placed between the particle's current location and the origin, and the particle relaxes asymptotically to rest at the origin ($D_x^\infty=0$), hence both trapping and cooling are achieved. 
If $b>1$ then these discrete dynamics become unstable and produce neither cooling nor trapping, as the trap is repeatedly placed further from the origin than the particle's current location.

To model the continuous version of this protocol and account for the fact that a real experimental apparatus does not respond instantaneously, we treat $D_x(t)$ as a dynamical variable that decays exponentially towards $x(t)$. Specifically, the system $(x,p)$ and parameter $D_x$ obey the equations of motion,
\begin{equation}
    \label{eq:classical_system}
    \begin{aligned}
        &\dot{x} = \frac{p}{m}, \\
        &\dot{p} = -m\omega^2 \left( x - bD_x \right), \\
        &\dot{D}_x = \gamma \left( x - D_x \right),
    \end{aligned}
\end{equation}
where $\gamma>0$ is a decay rate that characterizes the intrinsic lag of the experimental apparatus.
The value $\gamma^{-1}$ roughly corresponds to $\Delta t$ in the discrete case.

For $b<1$, the dynamics generated by Eq.~\eqref{eq:classical_system} relax to a stable fixed point at $p^\infty = x^\infty = D_x^\infty = 0$, thereby achieving both trapping and cooling.
For $b > 1$, Eq.~\eqref{eq:classical_system} produces unstable solutions in which both $D_x$ and $x$ accelerate at an exponential rate away from the origin.  At the critical value $b = 1$, the system relaxes to a solution given by $p^\infty = 0$ and $x^\infty = D_x^\infty = D_{x0} + \gamma p_0/m\omega^2$ (Eq.~\eqref{eq:asymp-cts}); in this situation the particle is cooled but not trapped at the origin.

When $b=1$, the relaxation timescale can be divided into two different regimes: a) if $\gamma < 2\omega$ the system is underdamped, and the relaxation rate is proportional to $\gamma$; and b) if $\gamma > 2\omega$ the system is overdamped, and when $\gamma \gg \omega$ the relaxation rate is given by $\omega^2/\gamma$, thus cooling occurs slowly.

We see that the discrete and the continuous protocols produce similar behavior for $b<1$, $b=1$ and $b>1$.
In the quantum case that we consider below, we use protocols analogous to the continuous protocol just presented, and the intuition on how to cool the system by manipulating the potential energy is similar. However, two additional ingredients appear in the quantum model: measurement backaction and measurement noise. 

\subsection{Quantum model - cooling protocols}\label{sec:cooling}
Consider a quantum harmonic oscillator of mass $m$ and characteristic frequency $\omega$ undergoing continuous measurement and feedback. Introducing dimensionless operators $\hat{p} \rightarrow \hat{p}\sqrt{\hbar m \omega}$ and $\hat{x} \rightarrow \hat{x}\sqrt{\hbar/m\omega}$, the Hamiltonian in the absence of feedback becomes 
\begin{equation}
    \label{eq:bare_hamiltonian}
    \hat{H}_0 = \frac{\hbar\omega}{2}\left( \hat{p}^2 + \hat{x}^2 \right).
\end{equation}
To implement feedback, we study three protocols similar to the one considered in the classical example. We imagine that position and/or momentum are continuously measured, and cooling is achieved by implementing a time-dependent Hamiltonian based on these measurements. The measurements correspond to Hermitian operators $\hat{A}_i \in \{ \hat{x},\hat{p} \}$, and feedback is applied through parameters $D_i \in \{ D_x,D_p \}$, see Eqs.~\eqref{eq:H1}-\eqref{eq:Hc}. As discussed in greater detail in the next section, these parameters evolve similarly to $D_x$ in Eq.~\eqref{eq:classical_system}.

Our first quantum protocol mimics the classical example. The position $\hat{A}_x = \hat{x}$ is measured, and feedback is applied through the parameter $D_x$ of the Hamiltonian
\begin{equation}
    \label{eq:H1}
    \hat{H}_x(D_x) = \frac{\hbar\omega}{2}\left[ \hat{p}^2 + \left( \hat{x}- bD_x \right)^2 \right].
\end{equation}
We will refer to this situation as Protocol X, which is illustrated in Fig.~\ref{fig:prot1_diagram}. As in the classical model of Sec.~\ref{sec:classical_model}, the parameter $b$ controls how we move the minimum of the potential in response to the particle's measured location. We will see that $b$ plays an important role in the trapping condition.

The second protocol generalizes the first by allowing both position and momentum measurements $\hat{A}_x = \hat{x}$ and $\hat{A}_p = \hat{p}$. Here the feedback Hamiltonian is 
\begin{equation}
    \label{eq:H2}
    \hat{H}_{xp}({\bf D}) = \frac{\hbar\omega}{2}\left[ \left( \hat{p}-bD_p \right)^2 + \left( \hat{x}-bD_x \right)^2 \right],
\end{equation}
where ${\bf D} = (D_x,D_p)$. We will refer to this protocol as Protocol XP.

Our third protocol is inspired by the cold damping protocol \cite{Courty2001,PhysRevLett.83.3174,PhysRevLett.122.223601,PhysRevLett.99.017201,PhysRevA.77.033804,Wilson2015,PhysRevA.72.043826,PhysRevLett.96.043003} that is widely used for cooling and trapping experiments. In the cold damping protocol, measurements are performed in the position basis $\hat{A}_x = \hat{x}$ and feedback is applied to create a force to oppose the direction of the movement $H_{\text{feedback}} \propto \dot D_x \hat{x}$. However, the formalism of Ref.~\cite{PhysRevLett.129.050401} requires a protocol that depends on $D_x$ and not on its time-derivative $\dot D_x$. Hence, we introduce a feedback Hamiltonian with a term proportional to $-\hat{p}D_x$:
\begin{equation}
    \label{eq:Hc}
    \hat{H}_c(D_x) = \frac{\hbar\omega}{2}\left[ (\hat{p}-\mu D_x)^2 + \hat{x}^2\right].
\end{equation}
A quadratic term $(\hbar\omega/2)\mu^2D_x^2$ term was added to give the feedback Hamiltonian a ground state energy of $\hbar\omega/2$, as with the other protocols. Even though this shift is feedback-dependent, it does not change the physical behavior of the system.
We will refer to this strategy as Protocol C, cross-feedback, to differentiate it from the usual cold damping scheme. The dimensionless constant $\mu$ regulates the force exerted on the particle.

\subsection{Quantum Fokker-Planck Master Equation}
We now describe how we model these protocols. While many studies of continuous measurement and feedback have focused on quantum Brownian models and stochastic equations \cite{PhysRevE.85.031110, Magazz__2018,PhysRevLett.97.130402,Tang_2020, PhysRevA.49.1350, doi:10.1098/rsta.2011.0531,PhysRevE.105.044137}, we will instead use the master equation derived recently in Ref.~\cite{PhysRevLett.129.050401}.

In Ref.~\cite{PhysRevLett.129.050401}, a continuous measurement is constructed in the limit of weak but frequent measurements modeled using Kraus operators \cite{doi:10.1080/00107510601101934}. Unlike idealized projective measurements that gain information in a single, instantaneous operation, continuous measurements gather limited information over a finite measurement time. This process is modeled using a Gaussian white noise variable $z(t)$ that represents the outcome of measuring a Hermitian operator $\hat{A}$.

Feedback is applied by varying the Hamiltonian based on the measurement outcome.  However, any real experimental apparatus has a finite bandwidth for information processing and feedback. We model this feature by writing the Hamiltonian as a function of a time-dependent parameter, $H=H(D(t))$, where $D(t)$ is a bandwidth-filtered version of the signal $z(t)$:
\begin{equation}
    \label{eq:ODE_D}
    \dot{D} = \gamma ( z - D ),
\end{equation}
(compare with Eq.~\eqref{eq:classical_system}). Equivalently,
    \begin{equation}
    \label{eq:filter_D}
    D(t) = \gamma \int_{-\infty}^t \dd s~e^{-\gamma(t-s)} z(s).
\end{equation}
$D(t)$ is treated as a classical, random variable.

These ingredients are captured by a Quantum Fokker-Planck master equation ~\cite{PhysRevLett.129.050401}, extended here to measurements of multiple observables (see Appendix~\ref{sec:A_multiple_measurements}):
\begin{equation}
\label{eq:QFPME}
    \begin{aligned}
        &\partial_t \hat{\rho}({\bf D},t) = \\
        &- \frac{i}{\hbar} \left[ \hat{H}({\bf D}),\hat{\rho}({\bf D},t) \right]  + \sum_j \lambda_j \mathcal{D}[\hat{A}_j]\hat{\rho}({\bf D},t)\\ 
        &+ \sum_j \left[ -\gamma_j\partial_{D_j} \mathcal{A}_j\hat{\rho}({\bf D},t)
            + \frac{\gamma_j^2}{8\lambda_j}\partial_{D_j}^2 \hat{\rho}({\bf D},t)
        \right],
    \end{aligned}
\end{equation}
The index $j$ runs over the measured observables $(\hat{A}_1, \hat{A}_2, \dotsc)$ and associated detector variables ${\bf D} = (D_1, D_2, \dotsc)$, and $\hat{\rho}({\bf D},t)$ is the joint state for the quantum system and the classical detector variables ${\bf D}$.

The first term on the right side of Eq.~\eqref{eq:QFPME} generates unitary evolution under $\hat{H}({\bf D})$. The second term represents the measurement backaction, and $\lambda_j$ denotes the strength of the $j$'th measurement. The superoperator $\mathcal{D}[\hat{c}]\hat{\rho} \equiv \hat{c} \hat{\rho} \hat{c}^\dagger - \frac{1}{2}\hat{c}^\dagger \hat{c} \hat{\rho} - \frac{1}{2}\hat{\rho} \hat{c}^\dagger \hat{c}$ is the dissipator, where $\hat{c}$ is an arbitrary operator. The third term, representing the system-detector coupling, describes the drift of each measurement outcome $D_j$ toward a value determined by the system state and the measured operator $\hat{A}_j$. Here, $\mathcal{A}_j\hat{\rho} \equiv \frac{1}{2}\{\hat{A}_j-D_j,\hat{\rho}\}$, where $\{\cdot,\cdot\}$ denotes the anti-commutator. The fourth term represents the diffusion of $D_j$ due to measurement noise, where $\gamma_j/8\lambda_j$ determines the magnitude of the noise. The constant $\gamma_j$ represents the $j$'th detector bandwidth, hence $\gamma_j^{-1}$ is the detector delay. See Ref.~\cite{PhysRevLett.129.050401} for the derivation and a more detailed discussion of the QFPME. 

Since $\hat{\rho}({\bf D},t)$ is a hybrid distribution over the combined spaces of the quantum system and the classical detector, expectation values are calculated jointly over these spaces:
\begin{equation}
\label{eq:ev_QD}
    \langle \hat{f}({\bf D}) \rangle(t) = \int \dd {\bf D} \Tr \left[ \hat{f}({\bf D}) \hat{\rho}({\bf D},t) \right] ,
\end{equation}
where $\hat{f}({\bf D})$ is a ${\bf D}$-dependent operator representing an observable of interest. We will focus on the evolution of such expectation values rather than attempting to solve  Eq.~\eqref{eq:QFPME} for $\hat{\rho}({\bf D},t)$.

\section{Results}\label{sec:results}
In the following sections, we apply Eq.~\eqref{eq:QFPME} to the protocols described by Eqs.~(\ref{eq:H1}-\ref{eq:Hc}).
In order to focus on the cooling and trapping produced by these protocols, we present the main results below, leaving technical details of their derivations to Appendices.

\subsection{Protocol X: position measurement}\label{sec:x-measurement}

Protocol X, defined by Eq.~\eqref{eq:H1} for $\hat{H}_x(D_x)$, involves continuous measurement of position. The joint state $\hat{\rho}(D_x,t)$ obeys (see Eq.~\eqref{eq:QFPME})
\begin{equation}
    \label{eq:QFPME_x}
    \begin{aligned}
        &\partial_t \hat{\rho} =  -\frac{i}{\hbar}\left[ \hat{H}_x, \hat{\rho} \right] + \lambda \mathcal{D}[\hat{x}]\hat{\rho} \\ &-\frac{\gamma}{2}\partial_{D_x} \{\hat{x}-D_x,\hat{\rho}\}
        + \frac{\gamma^2}{8\lambda}\partial_{D_x}^2 \hat{\rho}.
    \end{aligned}
\end{equation}
For simplicity, we omit the subscript $x$ in the parameters $\gamma$ and $\lambda$. This protocol is the quantum analog of the protocol discussed in Sec.~\ref{sec:classical_model}. Derivations of the results presented here are found in Appendix~\ref{sec:A-1measurement}.

\subsubsection{Energy and trapping}\label{sec:x-energy}
Using Eqs.~\eqref{eq:ev_QD} and \eqref{eq:QFPME_x}, we find that to solve for the expectation value of the feedback Hamiltonian $\hat{H}_x(D_x)$ we must solve a system of six coupled linear differential equations. The system can be simplified to three coupled equations in the case $b = 1$, as discussed in Appendix \ref{sec:A-1measurement}. From the fixed point of this set of equations, we find that the energy relaxes to
\begin{equation}
    \label{eq:energy-H1}
    \begin{aligned}
        \sev{\hat{H}_x}_\infty = &\frac{\hbar\omega}{2}\left[ \frac{\lambda}{b\gamma} + \frac{b\gamma}{4\lambda} + \frac{(2-b)\gamma\lambda}{2b\omega^2} \right].
    \end{aligned}
\end{equation}
A numerical verification of Eq.~\eqref{eq:energy-H1} for $b = 1$ is shown in Fig.~\ref{fig:numerics_H1}. We also find, from the same fixed point, that the position variance $\sigma_{x}^2 \equiv \sev{\hat{x}^2} - \sev{\hat{x}}^2$ relaxes to
\begin{equation}
    \label{eq:variance_x-H1}
    \begin{aligned}
        \sigma_{x,\infty}^2 =  &\frac{1}{2}\left[ \frac{\lambda}{b\gamma} + \frac{b\gamma}{4(1-b)\lambda} + \frac{\gamma\lambda}{b(1-b)\omega^2} \right].
    \end{aligned}
\end{equation}
This variance diverges when $b \rightarrow 1$, indicating that the protocol does not trap the particle at the origin. Intuitively, when $b=1$ there is no restoring force pushing the values of $\sev{x}$ and $\sev{D_x}$ toward the origin.
Rather, due to the underlying measurement noise, the particle and the detector together perform unbiased Brownian motion.

The timescale for the system to relax to its asymptotic state is determined from the eigenvalues of the matrix in Eq.~\eqref{eq:A_H1_matrix_form}. 
While we do not have a general, analytical expression for these eigenvalues, in two specific regimes the relaxation rate $\chi$ simplifies considerably: 
\begin{eqnarray}
\label{eq:protXregime1}
    &b = 1, \gamma < 2\omega &\quad\Rightarrow \chi = \gamma, \\
\label{eq:protXregime2}
    &\omega \gg \gamma &\quad\Rightarrow \chi = \gamma b.
\end{eqnarray}
If $b$ is small, the protocol is being applied weakly, and the relaxation is slow.

\subsubsection{Optimal regime and fine-tuning}
To optimize the protocol, we can choose values of $b$ that minimize either the energy or the position variance, depending on whether we prioritize cooling or trapping.

We minimize the asymptotic energy by setting
\begin{equation}
    \label{eq:opt-b-H1-energy}
    b = b_e = 2\lambda\sqrt{\frac{1}{\omega^2} + \frac{1}{\gamma^2}}.
\end{equation}
Note that this solution is valid only when $b_e < 1$.

The position variance is minimized by setting
\begin{equation}
    \label{eq:opt-b-H1-trap}
    b = b_v = \left(1+\frac{\gamma}{2\lambda}\sqrt{\frac{4\lambda^2+\omega^2}{\gamma^2+\omega^2}} \right)^{-1}.
\end{equation}
For $\omega \gg \gamma \gg \lambda$ and $b = b_v$ (note that in this regime $b_v\simeq b_e$), we obtain 
\begin{eqnarray}
    \sigma_{x,\infty}^2 &=& \frac{1}{2}\left( 1 + \frac{\lambda}{\gamma} + \frac{\gamma^2}{2\omega^2}\right) + \mathcal{O}(\gamma\lambda/\omega^2) \\
    \sev{\hat{H}_x}_\infty &=& \frac{\hbar\omega}{2}\left( 1 + \frac{2\lambda^2}{\gamma^2} + \frac{\gamma^2}{2\omega^2} \right) + \mathcal{O}(\gamma\lambda/\omega^2).
\end{eqnarray}
This result shows that trapping and cooling can be achieved simultaneously, to arbitrary accuracy. In this regime, the feedback is weak ($b\ll 1$) and the relaxation occurs slowly.

\subsection{Protocol XP: position and momentum measurements}\label{sec:xp-measurement}
In Protocol XP, the feedback uses weak measurements of both position and momentum.
The feedback Hamiltonian is thus parametrized by $D_x$ and $D_p$ (see Eq.~\eqref{eq:H2}).

For this protocol the QFPME for $\hat{\rho}({\bf D},t)$ is
\begin{equation}
    \label{eq:QFPME_xp}
    \begin{aligned}
        \partial_t \hat{\rho} = & -\frac{i}{\hbar}\left[ \hat{H}_{xp}, \hat{\rho} \right] + \lambda_x \mathcal{D}[\hat{x}]\hat{\rho} + \lambda_p \mathcal{D}[\hat{p}]\hat{\rho}\\
        &+ \gamma_x \mathcal{F}_x \hat{\rho} + \gamma_p \mathcal{F}_p \hat{\rho}, \\
    \end{aligned}
\end{equation}
where $\mathcal{F}_i \hat{\rho} = -(1/2)\partial_{D_i} \{ \hat{i} - D_i, \hat{\rho} \} + (\gamma_i/8\lambda_i) \partial_{D_i}^2 \hat{\rho}$ and $i = x,p$.
The detector bandwidths $\gamma_i$ and measurement strengths $\lambda_i$ characterize the measurement devices. For simplicity, we set $\lambda_x = \lambda_p = \lambda$ and $\gamma_x = \gamma_p = \gamma$.

\subsubsection{Energy and trapping}
\label{sec:xp-HS}

\begin{figure*}[!ht]
     \centering
     \begin{subfigure}[t]{0.32\textwidth}
         \centering
         \includegraphics[width=1\textwidth]{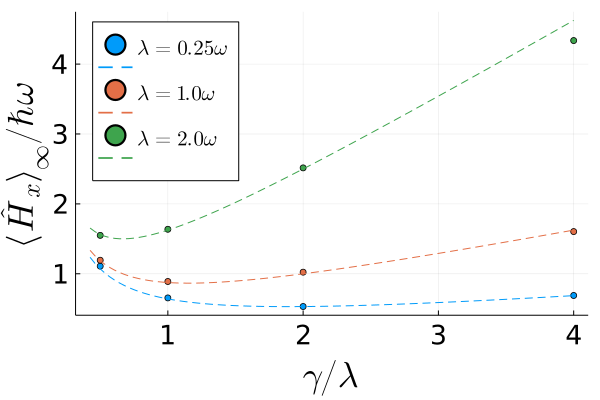}
         \caption{}
         \label{fig:numerics_H1}
     \end{subfigure}
     \begin{subfigure}[t]{0.32\textwidth}
         \centering
         \includegraphics[width=1\textwidth]{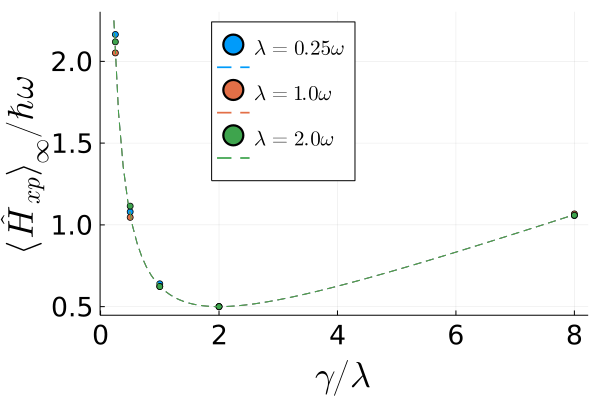}
         \caption{}
         \label{fig:numerics_H2}
     \end{subfigure}
     \begin{subfigure}[t]{0.32\textwidth}
         \centering
         \includegraphics[width=1\textwidth]{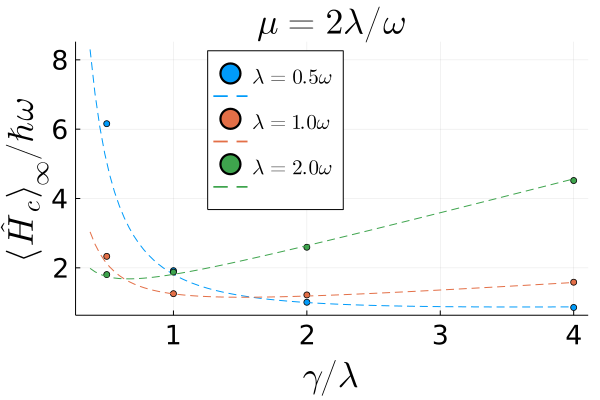}
         \caption{}
         \label{fig:numerics_H3}
     \end{subfigure}
     \caption{Comparison of analytical formulas (dashed lines) to numerical results (circles). We chose $b = 1$ for Protocols X and XP for simplicity. The average asymptotic energy was calculated using 1000 runs of the stochastic equations for (a) Protocol X for different values of measurement strength. (b) Protocol XP for different values of measurement strength. Here, we can see that the energy only depends on the ratio $\gamma / \lambda$. The non-monotonic energy behavior as a function of $\gamma/\lambda$ for both Protocols X and XP can be explained by the competition between the noise sources. See Sec.~\ref{sec:trajectory} for a detailed discussion. (c) Cross-feedback protocol, where we chose $\mu = 2\lambda / \omega$. Note that this plot illustrates that we need $\gamma \gg \omega \gg \lambda$ to achieve maximum cooling.}
         \label{fig:numerics_H}
\end{figure*}

As in Sec.~(\ref{sec:x-measurement}), we derive asymptotic solutions for the average energy $\sev{\hat{H}_{xp}}_\infty$ and position variance $\sigma_{x,\infty}^2$ (see Appendix~\ref{sec:A-2measurements} for details):
\begin{align}
    \label{eq:energy-H2}
    &\sev{\hat{H}_{xp}}_\infty = \frac{\hbar\omega}{2} \left[ \frac{\lambda}{b\gamma} + \frac{b\gamma}{4\lambda} + \frac{(1-b)\gamma\lambda}{2\omega^2} \right] \\
    \label{eq:variance_x-H2}
    &\begin{aligned}
        \sigma_{x,\infty}^2 =  &\frac{1}{2}\left[ \frac{\lambda}{b\gamma} + \frac{b\gamma}{4(1-b)\lambda} + \frac{\gamma\lambda}{b(1-b)\omega^2} \right].
    \end{aligned}
\end{align}
These solutions were derived from the fixed point of a system of four coupled differential equations. Note that the position variance is the same for Protocols X and XP.

As with Protocol X, the variance in position diverges when $b \rightarrow 1$. This occurs because $\sev{\hat{x}}_\infty = 0 = \sev{D_x}_\infty$ is no longer a stable fixed point, and the particle diffuses in space much like a Brownian particle. Figure ~\ref{fig:numerics_H2} presents a numerical verification of Eq.~\eqref{eq:energy-H2} for $b = 1$.

The relaxation rate to asymptotic values in Eqs.~\eqref{eq:energy-H2} and ~\eqref{eq:variance_x-H2} is given by the smallest eigenvalue of the $4\times 4$ matrix in Eq.~\eqref{eq:A_H2_matrix_form}:
\begin{equation}
    \label{eq:xp-b-frequency}
    \chi = \min{\mathfrak{Re}\left(\gamma \pm \frac{\Delta_\pm}{\sqrt{2}}\right)},
\end{equation}
with
\begin{equation}
    \begin{aligned}
        \Delta_\pm &= \sqrt{-\omega^2+\gamma^2 \pm  \Omega} \\
        \Omega &= \sqrt{\omega^4+2 \left(1-8b+8 b^2\right) \omega^2 \gamma ^2+\gamma^4}.
    \end{aligned}
\end{equation}
In two regimes, the relaxation rate $\chi$ simplifies as follows:
\begin{eqnarray}
    &b = 1 &\quad\Rightarrow \chi = 2\gamma, \\
    &\omega \gg \gamma &\quad\Rightarrow \chi = 2\gamma b.
\end{eqnarray}
Compare with Eqs.~\eqref{eq:protXregime1} and \eqref{eq:protXregime2}.
As with Protocol X, this protocol becomes slow as the feedback weakens in the limit $b \rightarrow 0$. Note that, using two measurements, the relaxation timescale is twice as fast as Protocol X.

\subsubsection{Optimal protocol}\label{sec:HXP_optimal}
Because the energy and variance solutions for Protocol XP (Eqs.~\eqref{eq:energy-H2}, \eqref{eq:variance_x-H2}) are similar to the ones found in Protocol X (Eqs.~\eqref{eq:energy-H1}, \eqref{eq:variance_x-H1}), the optimal parameter choices and the associated asymptotic position variance and energy values are the same for both protocols, up to $\mathcal{O}(\gamma\lambda/\omega^2)$.

The asymptotic energy for Protocol XP is smaller than the one achieved by Protocol X by a term $\hbar\omega\gamma\lambda/4\omega^2$. This is important because if one is not interested in trapping the particle, then the choices $b = 1$ and $\gamma = 2\lambda$ achieve ground-state cooling arbitrarily fast (see Appendix \ref{sec:A-2measurements}):
\begin{equation}
    \sev{\hat{H}_{xp}}_\infty = \frac{\hbar\omega}{2}.
\end{equation}

\subsection{Protocol C: Cross feedback}\label{sec:cold_damping}

Next, we consider the third protocol, with $\hat{H}_c(D_x)$ defined by Eq.~\eqref{eq:Hc}. Here, measurements performed on the particle's position are translated into feedback applied to its momentum, and the QFPME is 
\begin{equation}
    \label{eq:QFPME_Hc}
    \begin{aligned}
        &\partial_t \hat{\rho} =  -\frac{i}{\hbar}\left[ \hat{H}_c, \hat{\rho} \right] + \lambda \mathcal{D}[\hat{x}]\hat{\rho} + \gamma\mathcal{F}_x \hat{\rho}
    \end{aligned}
\end{equation}

\subsubsection{Energy and trapping}
The relevant quantities to solve for the asymptotic energy using the QFPME are determined by six coupled linear ODEs, see Appendix~\ref{sec:A-cross_feedback}.

Because of the lack of symmetries in the problem, we could not solve for the eigenvalues of the matrix in Eq.~\eqref{eq: Ml matrix}. 
As a result, for an arbitrary choice of parameters $(\omega,\lambda,\gamma,\mu)$, we do not know when the system converges asymptotically to a fixed value.
However, if the solution does converge, then the asymptotic energy with respect to the Hamiltonian in Eq.~\eqref{eq:Hc} is:
\begin{equation}
    \label{eq:energy-H3}
    \sev{\hat{H}_{c}}_\infty = \frac{\hbar\omega}{2}\left( \frac{\mu\omega}{4\lambda} + \frac{\lambda}{\mu\omega} + \frac{\lambda}{2\gamma} + \frac{\lambda\omega}{\gamma^2\mu} + \frac{\gamma\mu^2}{8\lambda} \right).
\end{equation}
We plot a numerical verification of Eq.~\eqref{eq:energy-H3} for $\mu = 2\lambda/\omega$ in Fig.~\ref{fig:numerics_H3}.

The position variance can also be obtained from the asymptotic solution, see Eq.~\eqref{eq:A_variance_position_Hc}:
\begin{equation}
    \label{eq:x-moved-cold-damping}
    \sigma_{x,\infty}^2 = \frac{1}{2}\left( \frac{\mu\omega}{4\lambda} + \frac{\lambda}{\mu\omega} + \frac{\lambda\omega}{\gamma^2\mu} \right).
\end{equation}
We see that this protocol is able to trap the particle (i.e.\ $\sigma_{x,\infty}^2$ is finite) for all positive values of $\mu$, in contrast with the other two protocols that required $b<1$ to achieve trapping.

\subsubsection{Optimal protocol}\label{sec:Prot3_optimal}
We can verify from Eq.~\eqref{eq:energy-H3} that no choice of parameters produces ground-state cooling. 
Finding the values of $\mu$, $\lambda$, $\gamma$ and $\omega$ that minimize $\sev{\hat{H}_{c}}_\infty$ is difficult, but if we assume $1 \gg \omega/\gamma \gg \lambda/\omega, \mu$ then the value of $\mu$ that minimizes the asymptotic energy is
\begin{equation}
    \label{eq:optimal_mu}
    \mu^* = 2\lambda/\omega \ll 1,
\end{equation}
leading to
\begin{equation}
    \sev{\hat{H}_c}_\infty =  \frac{\hbar\omega}{2}\left( 1 + \frac{\lambda\gamma}{2\omega^2}+ \frac{\omega^2}{2\gamma^2} \right) + \mathcal{O}(\lambda/\gamma)
\end{equation}

For small $\mu$, the system's relaxation timescale is $\mu \omega$, to $\mathcal{O}(\mu)$ (see Eq.~\eqref{eq:eigenvalues_cold_damping}), which under optimal cooling conditions is given by $\mu^*\omega = 2\lambda$.

As with the other protocols, the optimal condition for cooling is the same as the optimal condition for trapping. Here, using $\mu = 2\lambda/\omega$, we have 
\begin{equation}
\label{eq:variance-H3}
    \sigma_{x,\infty}^2 = \frac{1}{2}\left( 1 + \frac{\omega^2}{2\gamma^2} \right)
\end{equation}
in dimensionless units.

\subsection{Coupling to a thermal 
bath}\label{sec:thermal-bath}

The above calculations were performed for a closed quantum system under continuous measurement and feedback.  We now consider the case in which the system is open. Specifically, it is coupled to a  thermal environment as it simultaneously undergoes measurement and feedback.  The environment's presence is modeled by adding Lindblad jump terms,
\begin{equation}
    \label{eq:bath_terms}
    \begin{aligned}
        \Gamma(\bar{n}+1)\mathcal{D}[\hat{a}_{i}]\hat{\rho}({\bf D}) + \Gamma\bar{n}\mathcal{D}[\hat{a}_{i}^\dagger]\hat{\rho}({\bf D}),
    \end{aligned}
\end{equation}
to the QFPME for each protocol.
Here $\Gamma$ denotes the system-bath coupling strength, and $\bar{n} = \left( e^{\beta \hbar \omega} -1 \right)^{-1}$ controls the asymptotic average energy the system would relax to, in the absence of feedback control:
\begin{equation}
    \label{eq:thermal_energy}
    U_{\text{th}}=\hbar \omega (\bar{n}+1/2).
\end{equation}

The operators $\hat{a}_{i}$ ($\hat{a}_{i}^\dagger$) with $i=\{x,xp,c\}$ are the annihilation (creation) operators:
\begin{align}
\label{eq:bath_a1}
    &\hat{a}_{x} = \sqrt{1/2}\left( \hat{x} - D_x +i \hat{p} \right), \\
    \label{eq:bath_a2}
    &\hat{a}_{xp} = \sqrt{1/2}\left( \hat{x} - D_x +i \hat{p} - iD_p \right), \\
    \label{eq:bath_a3}
    &\hat{a}_{c} = \sqrt{1/2}\left( \hat{x} +i \hat{p} -i\mu D_x \right).
\end{align}
Here, we focus on Protocols X and XP when $b = 1$, as the general case $b \neq 1$ is cumbersome and not particularly insightful.

The operators $\{\hat{a}_i\}$ satisfy $[\hat{H}_i, \hat{a}_{i}] = -\hbar \omega \hat{a}_{i}$ for $i\in\{x,xp,c\}$, hence the effect of the bath is to create and destroy single quanta of energy in the measurement-outcome-dependent energy basis.

Let us assume the following constraints for each protocol: a) Protocol X, $\omega \gg \gamma, \lambda, \Gamma$; b) Protocol XP, no constraints; c) Protocol C, $\gamma \gg \omega \gg \mu\gamma, \lambda\gamma/\omega,\Gamma\gamma/\omega$. We can then write the asymptotic energy as a weighted average,
\begin{equation}
    \label{eq:average_thermal_average}
    \sev{\hat{H}_i}_{\Gamma} = \frac{\Gamma_i \sev{\hat{H}_i}_{\infty} + \Gamma U_{\text{th}}}{\Gamma_i + \Gamma},
\end{equation}
where $i=\{ x,xp,c \}$. The weight factors $\Gamma_i$ are:
\begin{align}
    &\Gamma_x = \gamma, \\
    &\Gamma_{xp} = 2\gamma, \\
    &\Gamma_c = \mu\omega,
\end{align}

Equation~\eqref{eq:average_thermal_average} expresses the asymptotic energy as a weighted average of the feedback energy and the thermal energy, where the weights are the protocols' relaxation rates (close to their optimal conditions) and the thermal bath's relaxation rate. 
This weighted average reflects a competition between the bath, which tries to bring the system to thermal equilibrium, and the measurement and feedback protocol, which tries to minimize the system's energy.
In the absence of measurement and feedback, $\Gamma$ represents the relaxation rate to thermal equilibrium.
When measurement and feedback are present, but $\gamma,\lambda, \mu\omega \ll \Gamma$, then the effects of the bath are dominant and the system's asymptotic energy is close to the thermal equilibrium energy. Conversely, if $\gamma,\lambda, \mu\omega \gg \Gamma$, then the system is stabilized near the solution given by the asymptotic average energy of a closed system. Thus, thermal coupling imposes an additional requirement, namely $\gamma,\lambda,\mu\omega \gg \Gamma$, that must be satisfied to maximize cooling.

\section{Trajectory simulation}\label{sec:trajectory}

\begin{figure*}[!ht]
    \begin{subfigure}[t]{0.46\textwidth}
        \includegraphics[width=1\textwidth]{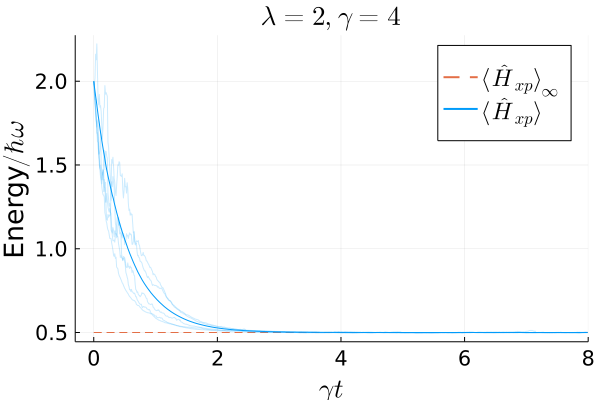}
        \caption{}
        \label{fig:trajectory_H2}
    \end{subfigure}
    \begin{subfigure}[t]{0.46\textwidth}
        \includegraphics[width=1\textwidth]{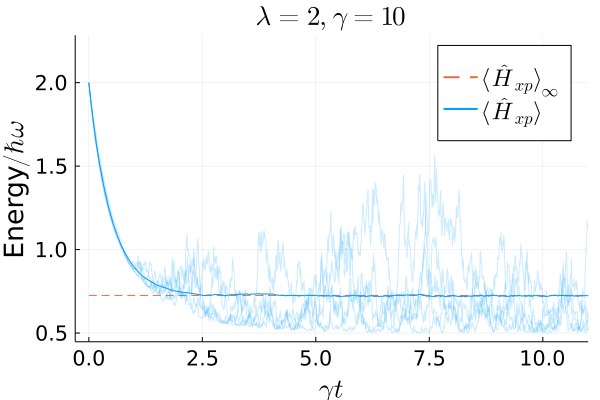}
        \caption{}
        \label{fig:ensemble_H2}
    \end{subfigure}
    \caption{Single trajectory simulation for Protocol XP ($b = 1$). Here, we plot 5 trajectories (light blue lines) and the average energy for 2000 trajectories (blue line). The trajectory energy, in general, only converges with the analytical result (dashed line) in the ensemble sense. See Sec.~\ref{sec:trajectory} for simulation details. (a) Single trajectory simulation for Protocol XP with ground state cooling ($\gamma=2\lambda$). The energy initially evolves stochastically, but after the variances have decayed to their asymptotic values, the energy evolution becomes deterministic, producing ground-state cooling at the trajectory level. (b) Here, the protocol applied is not at the ground-state condition ($\gamma \neq 2\lambda$). Note that the individual trajectories don't necessarily converge to a fixed energy value, but the ensemble average converges to the analytical result derived in Eq.~\eqref{eq:energy-H2}.}
    \label{fig:trajectory_simulation}
\end{figure*}

Here, we describe numerical simulations of stochastic trajectories that were used both to test the analytical results of Sec.~\ref{sec:results}, and to build trajectory-level intuition about the ground state cooling described in Sec.~\ref{sec:HXP_optimal}. The quantities described in this section are the noisy counterparts of the quantities discussed in previous sections.

Our starting point is the Belavkin equation \cite{10.1007/BFb0041197,BELAVKIN1989355,ABarchielli_1991,BELAVKIN1992171,PhysRevA.69.022109} coupled with the evolution of the measurement outcome:
\begin{align}
    \label{eq:belavkin}
    &\begin{aligned}
        d\hat{\rho} =& -\frac{i}{\hbar}[\hat{H}({\bf D}),\hat{\rho}]dt \, \\
        & + \lambda_x \mathcal{D}[\hat{x}]\hat{\rho} dt + \sqrt{\lambda_x} dW_x \{ \hat{x}-\langle \hat{x} \rangle, \hat{\rho} \}\\
        & + \lambda_p \mathcal{D}[\hat{p}]\hat{\rho} dt + \sqrt{\lambda_p} dW_p \{ \hat{p}-\langle \hat{p} \rangle, \hat{\rho} \},
    \end{aligned}\\
    \label{eq:belavkin_signal}
    &dD_i = \gamma_i(\langle \hat{i} \rangle-D_i) + \frac{\gamma_i}{\sqrt{4\lambda_i}}dW_i, \quad i=x,p.
\end{align}
The Wiener increments $dW_i$ are the same for both equations, and they satisfy $dW_i^2 = dt$ and $dW_x dW_p = 0$ \cite{gardiner1985handbook}.

Equations \eqref{eq:belavkin} and \eqref{eq:belavkin_signal} describe the coupled, stochastic dynamics of the system $\hat\rho$ and measurement outcomes $D_i$. Equation \eqref{eq:belavkin} is the Belavkin equation, which generalizes the Schrödinger equation to continuously monitored systems. The Belavkin equation relates the infinitesimal stochastic evolution of the density matrix to the unitary evolution, backaction, and random update based on the stochastic outcome of the measurement.  Equation \eqref{eq:belavkin_signal} describes the noisy relaxation of the detector variables $D_x$ and $D_p$ toward the average position and momentum $\langle \hat{x} \rangle$ and $\langle \hat{p} \rangle$.

Together, Eqs. \eqref{eq:belavkin} and \eqref{eq:belavkin_signal} are the single-trajectory counterparts of the QFPME, Eq.~\eqref{eq:QFPME}, which describes the combined evolution at the ensemble level. Equation \eqref{eq:belavkin} gives rise to the unitary part and the backaction terms in Eq.~\eqref{eq:QFPME} while Eq.~\eqref{eq:belavkin_signal} generates the Fokker-Planck component of the QFPME.

In Eq. \eqref{eq:belavkin}, the system's Hamiltonian has the general quadratic form
\begin{equation}
    \label{eq:general_H}
    \hat{H}({\bf D}) = \frac{\hbar \omega}{2}\left[ \left( \hat{p} - b_x D_x - b_p D_p \right)^2 + \left( \hat{x} - c_x D_x - c_p D_p \right)^2 \right],
\end{equation}
which can reproduce all the protocols introduced and investigated in Sec.~\ref{sec:model} and \ref{sec:results}.

Under Eq.~\eqref{eq:belavkin}, the evolution of $\langle \hat{x} \rangle$, $\langle \hat{p} \rangle$ and higher cumulants (such as variances) is described by a set of infinitely many coupled equations. As discussed in Appendix \ref{sec:A_trajectory}, if we assume an initial Gaussian wave function for the state of the system \cite{PhysRevA.57.2301,PhysRevA.60.2700,doi:10.1080/00107510601101934}, then the system's state remains Gaussian at all times.  The system's evolution is then uniquely defined by the evolution of the first two sets of cumulants, allowing us to write a closed set of stochastic differential equations (SDE's) for these cumulants and the detector variables $D_x$ and $D_p$; see Eq.~\eqref{eq:SDE-belavkin-general}.  We have numerically simulated trajectories evolving under these coupled SDE's, and by averaging over these trajectories, we have calculated the asymptotic energy values $\langle H\rangle_\infty$ for various protocols. Figure \ref{fig:numerics_H} shows good agreement between these numerically obtained values and the analytical results derived in Sec.~\ref{sec:results}.

The trajectory approach also provides insight into how Protocol XP achieves ground-state cooling for $b=1$ and $\gamma=2\lambda$ in Eq.\ \eqref{eq:energy-H2}. Using $b_x = c_p = 0$ and $b_p = c_x = 1$ in Eq.~\eqref{eq:general_H}, when we choose $\lambda_x = \lambda_p = \lambda$ the system evolves asymptotically to a symmetric Gaussian with variances (see Eq.\ \eqref{eq:symmetridAsymptoticGaussian})
\begin{equation}
\label{eq:asympVars}
    (V_x)_\infty = (V_p)_\infty = \frac{1}{2} .
\end{equation}
Note that these quantities differ from the previously discussed position variance. Here, they refer to the actual wave function position variance at the trajectory level. Conversely, the results shown before refer to the ensemble average position variance over many realizations of the experiment.

After the variances decay, the stochastic equation for the average energy $d\langle \hat{H}_{xp} \rangle = d\Tr [\hat{H}_{xp} \hat{\rho}]$, using Eq.~\eqref{eq:belavkin}, can be simplified using $\gamma_x = \gamma_p = \gamma$:
\begin{equation}
    \label{eq:dH-opt-protocol}
    \begin{aligned}
        d\langle \hat{H}_{xp} \rangle = &2\gamma\left( \sev{\hat{H}_{xp}}_\infty - \langle \hat{H}_{xp}\rangle \right)dt \\
        &+\hbar\omega\left( \langle \hat{x} \rangle - D_x \right) \left( 1 - \frac{\gamma}{2\lambda} \right) \sqrt{\lambda} dW_x \\
        &+\hbar\omega\left( \langle \hat{p} \rangle - D_p \right) \left( 1 - \frac{\gamma}{2\lambda} \right) \sqrt{\lambda} dW_p.
    \end{aligned}
\end{equation}
When $\gamma = 2\lambda$, the stochastic terms vanish and Eq.~\eqref{eq:dH-opt-protocol} reduces to exponential cooling to the ground state. See Fig.~\ref{fig:trajectory_simulation} for a comparison of how the energy evolves for a) $\gamma = 2\lambda$ and b) $\gamma \neq 2\gamma$. 

It is also instructive to look at the SDE's for $\langle\hat x\rangle$, $\langle\hat p\rangle$, $D_x$ and $D_p$ (see Eq.\ \eqref{eq:SDE-belavkin-general}, with parameters set as in the previous paragraph), which combine to give
\begin{equation}
\label{eq:XP}
    \begin{aligned}
        dX &= (-\gamma X + \omega P) \, dt + \left( \sqrt{\lambda} - \frac{\gamma}{\sqrt{4\lambda}} \right) dW_x, \\
        dP &= (-\gamma P - \omega X) \, dt + \left( \sqrt{\lambda} - \frac{\gamma}{\sqrt{4\lambda}} \right) dW_p,
    \end{aligned}
\end{equation}
for the quantities $X = \langle\hat x\rangle-D_x$ and $P = \langle\hat p\rangle-D_p$. The coefficient multiplying the Wiener increments $dW_x$ and $dW_p$ reflects a trade-off between contributions from the noise on the system's dynamics due to measurements, and the noise on the detector outcome.  The former is proportional to $\sqrt{\lambda}$ (Eq.\ \eqref{eq:belavkin}) reflecting the fact that increased measurement backaction leads to increased noise; the latter is proportional to $\gamma / \sqrt{4\lambda}$ (Eq.\ \eqref{eq:belavkin_signal}), reflecting both the detector bandwidth $\gamma$, and the fact that the increased measurement strength $\lambda$ leads to more precise (less noisy) measurement outcomes.  When $\gamma=2\lambda$ these contributions cancel in Eq.~\eqref{eq:XP}, and the resulting deterministic equations have the asymptotic solution
\begin{equation}
\label{eq:asympMeans}
    X_\infty = 0 \quad,\quad P_\infty = 0 .
\end{equation}
Eqs.\ \eqref{eq:asympVars} and \eqref{eq:asympMeans} describe the ground state of a harmonic oscillator whose minimum is located at the origin.

We emphasize that although in this section we assumed an initial Gaussian state for the system, the derivations in previous sections did not impose this restriction: the asymptotic solutions derived from the QFPME are independent of initial conditions.

\renewcommand{\arraystretch}{1.5}
\begin{table*}[!t]
\begin{tabular}{c|c|c|c|c}
Protocol & Regime & Timescale & Energy & Position Variance \\ \hline
X & $\omega \gg \gamma \gg \lambda$ & $\gamma b$ & $\frac{\hbar\omega}{2}\left( \frac{\lambda}{b\gamma} + \frac{b\gamma}{4\lambda} \right)$ & $\frac{1}{2}\left( \frac{\lambda}{b\gamma} + \frac{b\gamma}{4(1-b)\lambda} \right)$ \\ \hline
\multirow{2}{*}{XP} & $\gamma = 2\lambda, \quad b = 1$ & $2\gamma$ & $\frac{\hbar\omega}{2}$ & $\infty$ \\ 
& $\omega \gg \gamma \gg \lambda$ & $2\gamma b$ & $\frac{\hbar\omega}{2}\left( \frac{\lambda}{b\gamma} + \frac{b\gamma}{4\lambda} \right)$ & $\frac{1}{2}\left( \frac{\lambda}{b\gamma} + \frac{b\gamma}{4(1-b)\lambda} \right)$ \\ \hline
C & $\gamma \gg \omega \gg \lambda \gamma/\omega$ & $\mu\omega$ & $\frac{\hbar\omega}{2}\left( \frac{\mu\omega}{4\lambda} + \frac{\lambda}{\mu\omega} \right)$ & $\frac{1}{2}\left( \frac{\mu\omega}{4\lambda} + \frac{\lambda}{\mu\omega}\right)$

\end{tabular}
\caption{Performance of different protocols under optimal conditions. The column Regime denotes the hierarchy resulting in optimal conditions. The column Timescale shows the relaxation rate to the asymptotic value. The column Energy represents the asymptotic energy achieved. The column Position Variance shows the variance obtained after trapping the particle.  Note the special condition for ground-state cooling that can only be achieved by Protocol XP. The optimal choice for parameters to achieve near ground-state cooling are $b = 2\lambda/\gamma$ and $\mu = 2\lambda/\omega$.}
\label{tab:summary-results}
\end{table*}

\section{Discussion}\label{sec:discussion}

The results discussed in previous sections allow us to compare the three protocols on an equal footing. Each protocol has strengths and weaknesses that will affect the protocol choice for particular implementations. The level of control in the experiment plays a vital role in determining the success of this approach; the measurement operators and the range of values of the parameters are the primary factors in deciding which protocol to use.

The optimal protocols discussed in the previous sections can be summarized in Table~\ref{tab:summary-results}. All protocols can trap and cool down the particle for suitable choices of parameters. Protocol XP can quickly cool the particle to the ground state if no trapping is necessary. Protocol C can trap the particle close to the minimum uncertainty with a quadratic correction, as opposed to a linear correction from Protocols X and XP (compare Eq.~\eqref{eq:x-moved-cold-damping} to Eq.~\eqref{eq:variance_x-H1} and Eq~\eqref{eq:variance_x-H2}, in their optimal conditions).

The physical implementation and range of parameters are essential to decide which protocol to use. Note that the frequency is the largest parameter for protocols X and XP, while for protocol C, the bandwidth is the largest parameter. This can play an essential role in which protocols can be implemented. For Protocol X, measurement and feedback are performed in the position basis. For Protocol XP, we require both measurement and feedback in both position and momentum, and Protocol C requires measurement in position and feedback in momentum. In experiments, the choice of protocols can also be affected by which observables can be measured.

Overall, if we measure only a single variable (position or momentum), then Protocol C outperforms Protocol X for trapping; however, a larger separation of timescales is necessary $\gamma \gg \omega \gg \lambda\gamma/\omega$. If position and momentum measurements are feasible and there is no need for trapping, then Protocol XP gives the best result for fast ground state cooling. In all protocols, if the system is not isolated from a thermal environment, the cooling rate must be larger than the thermal relaxation rate to mitigate the heating effects of the thermal bath.

\section{Conclusion}\label{sec:conclusion}
We have studied cooling protocols using a recently derived framework for continuous measurement and feedback. We applied the Quantum Fokker-Planck Master equation \cite{PhysRevLett.129.050401} to a quantum harmonic oscillator that serves as a simple but relevant system for ground state cooling techniques. Unlike previous studies on measured harmonic oscillators \cite{PhysRevE.85.031110,PhysRevE.105.044137}, our model includes a finite detector bandwidth, which has proven to be a central ingredient for achieving ground state cooling.

We focused our study on three different measurement schemes: position measurement (X), position and momentum measurement (XP), and the cross-feedback protocol (C) which is related to cold damping. We showed that Protocol XP yields better results for the cooling task, and cooling can be done arbitrarily fast compared to a single measurement in position. Also, no separation of timescale needs to be satisfied (other than $\gamma \gg \Gamma$ to avoid thermal excitation).

Our main results are analytical expressions for the asymptotic energy achieved by these schemes and the emergence of optimum parameters to trap and cool the system. Feedback schemes are usually complex to solve; thus, our framework uncovers some techniques and tools that could be employed to solve more complicated systems. Future work will include an in-depth study of the thermodynamics of the QFPME, more general feedback protocols, and possible experimental verification of results.

Another interesting avenue for investigation is the squeezing effect that can happen due to continuous feedback \cite{PhysRevA.49.1350}. Although this question is out of the scope of this work, it can help answer the question of which interesting quantum states can be prepared using continuous feedback.

\acknowledgments
We thank Debankur Bhattacharyya for fruitful discussions. This research was supported by FQXi Grant No. FQXi- IAF19-07 from the Foundational Questions Institute Fund, a donor advised fund of Silicon Valley Community Foundation.  G.D.S and C.J. additionally acknowledge support from the John Templeton Foundation (award no. 62422). P. B. also acknowledges funding from the European Research Council (Consolidator grant ‘Cocoquest’ No. 101043705). B.A.A. was supported by the Swedish
Research Council, Grant No. 2018-03921. P.P.P. acknowledges funding from the Swiss National Science Foundation (Eccellenza Professorial Fellowship PCEFP2\_194268).

\bibliography{references}

\clearpage
\newpage
\widetext
\begin{center}
	\textbf{\large Appendix: Quantum Harmonic Oscillator under measurement and feedback}
\end{center}
\setcounter{equation}{0}
\setcounter{figure}{0}
\setcounter{table}{0}
\setcounter{page}{1}
\setcounter{section}{0}
\makeatletter
\renewcommand{\theequation}{A\arabic{equation}}
\renewcommand{\thefigure}{A\arabic{figure}}

This appendix provides technical details and sketches of derivations of the results presented in the main text.

\section{Feedback cooling in a classical oscillator}\label{sec:A_classical_model}
Here, we present details of the analysis of the classical model discussed in Sec.~\ref{sec:classical_model} of the main text.

\subsection{Discrete time feedback}
Letting $\left( x_k, p_k \right) \equiv \left( x(k\Delta t), p(k\Delta t) \right)$ denote the particle's state at stroboscopic times $k\Delta t$, $k = 0, 1, 2, \dotsc$, the evolution from one step to the next is given by the linear map
\begin{equation}
    \begin{pmatrix}
        x_{k+1} \\ p_{k+1}
    \end{pmatrix} = 
    \begin{pmatrix}
        b + (1-b)\cos(\omega\Delta t) & \sin(\omega\Delta t) / m\omega \\ 
        -m\omega(1-b)\sin(\omega\Delta t) & \cos(\omega\Delta t)
    \end{pmatrix}
    \begin{pmatrix}
        x_k \\ p_k
    \end{pmatrix}.
\end{equation}
We assume $\Delta t$ is not an integer multiple of $\pi/\omega$.
The particle's long-time evolution is determined by the eigenvalues of the above matrix.  When $b<1$, the magnitudes of both eigenvalues are smaller than unity, and the state vector $(x_k,p_k)$ decays to the unique stationary state $(0,0)$ as $k\rightarrow\infty$, i.e.\ the particle is cooled and trapped at the origin.
When $b>1$, one eigenvalue's magnitude exceeds unity, hence for generic initial conditions, both $x_k$ and $p_k$ diverge exponentially with $k$.
Finally, when $b=1$, the matrix is easily diagonalized, and we obtain
\begin{equation}
\label{eq:asymp-discrete}
     x_k \rightarrow x_0 + \frac{\sin(\omega\Delta t)}{1-\cos(\omega\Delta t)} \frac{p_0}{m\omega}
    \quad,\quad
    p_k \rightarrow 0
\end{equation}
as $k\rightarrow\infty$, thus the particle is cooled but not trapped at the origin.

\subsection{Continuous feedback}
Defining $\zeta \equiv (x,p,D_x)^T$ and writing Eq.~\eqref{eq:classical_system} as $\dot\zeta = M \zeta$, we see that the evolution of the particle's state $(x,p)$ and its  measured position $D_x$ is governed by the matrix
\begin{eqnarray}
M = 
    \begin{pmatrix}
        0 & 1/m & 0 \\
        -m\omega^2 & 0 & m\omega^2 b \\
        \gamma & 0 & -\gamma
    \end{pmatrix} .
\end{eqnarray}
When $b<1$, all eigenvalues of $M$ have strictly negative real parts, which implies that the system evolves asymptotically to the state $\zeta^\infty = (0,0,0)$,  corresponding to cooling and trapping at the origin.
When $b>1$, one of $M$'s eigenvalues has a positive real part, producing exponential divergence away from the origin; in this case, neither cooling nor trapping is achieved.

When $b = 1$, the eigenvalues of $M$ are $0$ and $\pm i\Omega$, where $\Omega = \sqrt{\omega^2 - \gamma^2/4}$.  The explicit solution of Eq.~\eqref{eq:classical_system}, for initial conditions $(x_0,p_0,D_{x0})$, is then:
\begin{align}
    &p(t) = p_0 e^{-\gamma t /2} \cos(\Omega t) + \frac{1}{\Omega} \left[ \frac{\gamma}{2}p_0 - m\omega^2(x_0 - D_{x0}) \right] e^{-\gamma t /2} \sin(\Omega t), \\
    &x(t) = x_0 + \left[ \frac{\gamma p_0}{m\omega^2} - (x_0 - D_{x0}) \right]\left( 1 - e^{-\gamma t/2}\cos(\Omega t) \right) + \left[ \frac{p_0}{\Omega m \omega^2} (\Omega^2 - \gamma^2/4) + \frac{\gamma}{2\Omega}(x_0-D_{x0}) \right] e^{-\gamma t /2} \sin(\Omega t), \\
    &D_x(t) = D_{x0} + \frac{\gamma p_0}{m\omega^2} \left( 1 - e^{-\gamma t/2} \cos(\Omega t) \right) + \left[ -\frac{\gamma p_0}{2m\omega^2} + (x_0-D_{x0}) \right]\frac{\gamma}{\Omega} e^{-\gamma t/2} \sin(\Omega t).
\end{align}
Asymptotically with time, this solution gives
\begin{equation}
\label{eq:asymp-cts}
    x(t), \, D_x(t) \rightarrow D_{x0} + \frac{\gamma p_0}{m \omega^2}
    \quad,\quad
    p(t) \rightarrow 0 .
\end{equation}
Thus, the particle ends at rest at the bottom of the trap, but in general the trap is not located at the origin.

\section{QFPME with multiple measurements}\label{sec:A_multiple_measurements}
Here, we outline the steps used to derive the QFPME for two continuous measurements, Eq.~\eqref{eq:QFPME}.  Similar steps hold for any finite number of measurements. This derivation is based on the stochastic calculus section of Supplemental Material on Ref.~\cite{PhysRevLett.129.050401} and Appendix B of Ref.~\cite{wiseman_milburn_2009}.

We start with the Belavkin equation for two measurements \cite{PhysRevA.69.022109,doi:10.1080/00107510601101934}
\begin{equation}
    \begin{aligned}
    \label{eq:A_Belavkin_2}
        d\hat{\rho}_c = &-\frac{i}{\hbar}\left[ \hat{H}, \hat{\rho}_c \right] dt +\lambda_1 \mathcal{D}[\hat{A}_1]\hat{\rho}_c dt + \lambda_2 \mathcal{D}[\hat{A}_2]\hat{\rho}_c dt + \\
        &\sqrt{\lambda_1} \{ \hat{A}_1 - \langle\hat{A}_1\rangle, \hat{\rho}_c \} dW_1 + \sqrt{\lambda_2} \{ \hat{A}_2 - \langle\hat{A}_2\rangle, \hat{\rho}_c \} dW_2,
    \end{aligned}
\end{equation}
where $\hat{\rho}_c$ is the density matrix conditioned on a particular stochastic trajectory for the outcome variables $D_{1,2}^'$. The operators $\hat{A}_{1,2}$ are measurement operators, $\lambda_{1,2}$ are the corresponding measurement strengths, and $\langle\hat{A}_{1,2}\rangle = \Tr(\hat{A}_{1,2}\hat{\rho}_c)$ are conditioned expected values. The stochastic variables $dW_{1,2}$ are Wiener processes that satisfy $dW_{i}dW_{j} = \delta_{ij}dt$ and $\mathrm{E}[dW_{1,2}]= 0$. For brevity, we have suppressed the dependence of the Hamiltonian on the measurement outcomes $\hat{H}=\hat{H}(D_1^',D_2^')$ for the sake of notation. This equation is coupled to the evolution of the filtered measurement outcomes
\begin{equation}
\label{eq:A_evolution_dD}
    dD_i^' = \gamma_i \left( \langle\hat{A}_i\rangle - D_i^' \right)dt + \frac{\gamma_i}{\sqrt{4\lambda_i}}dW_i, \quad i = 1,2,
\end{equation}
where $\gamma_{1,2}$ are the detector bandwidths.

We are interested in the joint distribution of the system density matrix and detector outcomes $\hat{\rho}(D_1,D_2) = \mathrm{E}_{W_1,W_2}[\hat{\rho}_c \delta(D_1 - D_1^') \delta(D_2 - D_2^')]$, where the average $\mathrm{E}_{W_1,W_2}[\cdot]$ is taken over the noisy variables $W_{1,2}$. To obtain the equation of motion for this object, we use Ito's rules
\begin{subequations}
    \begin{align}
        &d(fg) = fdg + df g + df dg, \\
        &df(y) = f'(y) dy + \frac{1}{2} f''(y) (dy)^2,
    \end{align}
\end{subequations}
with $y(t)$ evolving under a stochastic process $dy = a(y) dt + b(y) dW$. Using these rules, we find that 
\begin{equation}
    \begin{aligned}
        d[\hat{\rho}_c \delta(D_1 - D_1^') \delta(D_2 - D_2^')] = &~d\hat{\rho}_c \delta(D_1^' - D_1) \delta(D_2^' - D_2) \\
        &+ \hat{\rho}_c d\delta(D_1^' - D_1) \delta(D_2^' - D_2) + \hat{\rho}_c\delta(D_1^' - D_1) d\delta(D_2^' - D_2) \\
        &+ d\hat{\rho}_c d\delta(D_1^' - D_1) \delta(D_2^' - D_2) + d\hat{\rho}_c \delta(D_1^' - D_1) d\delta(D_2^' - D_2).
    \end{aligned}
\end{equation}
Note that the term $\hat{\rho}_c d\delta(D_1^' - D_1) d\delta(D_2^' - D_2)$ can be suppressed as it gives no contribution $\mathcal{O}(dt)$ because $\mathrm{E}[dW_1 dW_2] = 0$. Now, one can use the following identities
\begin{subequations}
    \begin{align}
        &\frac{\partial}{\partial_{D_i^'}}\delta(D_i^'-D_i) = -\frac{\partial}{\partial_{D_i}}\delta(D_i'-D_i), \\
        &\left[ \frac{\partial}{\partial_{D_i^'}}\delta(D_i^'-D_i) \right] f(D_i^') = \frac{\partial}{\partial_{D_i^'}}\left[ \delta(D_i^'-D_i) f(D_i) \right],
    \end{align}
\end{subequations}
for an arbitrary smooth function $f(D_i^')$. Combining all these equations and taking the ensemble average of measurement outcomes, we obtain the QFPME with multiple measurements appearing in Eq.~\eqref{eq:QFPME}.

\section{Protocol X: Position Measurement} \label{sec:A-1measurement}
To solve for the asymptotic expectation value of $\hat{H}_x({\bf D})$, we first use Eq.~\eqref{eq:QFPME_x} to write a set of coupled differential equations. In matrix form, these equations are:
\begin{equation}
    \label{eq:A_H1_matrix_form}
    \partial_t\begin{pmatrix}
        \nu_1 \\ \nu_2 \\ \nu_3 \\ \nu_4 \\ \nu_5 \\ \nu_6
    \end{pmatrix}
    = \begin{pmatrix}
        -2\gamma b & 0 & \frac{\omega}{2} &  -\gamma b(b-1) & 0 & 0 \\
        0 & 0 & -\frac{\omega}{2} & 0 & 0 & 0 \\ 
        -4\omega & 4\omega & -\gamma b & 0 & 2\gamma b(1-b) & 0 \\
        2\gamma & 0 & 0 & -\gamma & \omega & \gamma b(1-b) \\ 
        0 & 0 & \frac{\gamma}{2} & -\omega & -\gamma(1-b) & 0 \\
        0 & 0 & 0 & 2\gamma & 0 & -2\gamma(1-b)
    \end{pmatrix}
    \begin{pmatrix}
        \nu_1 \\ \nu_2 \\ \nu_3 \\ \nu_4 \\ \nu_5 \\ \nu_6
    \end{pmatrix}
    + \begin{pmatrix}
        \frac{b^2\gamma^2}{8\lambda} \\ \frac{\lambda}{2} \\ 0 \\ -\frac{b\gamma^2}{4\lambda} \\ 0 \\ \frac{\gamma^2}{4\lambda}
    \end{pmatrix},
\end{equation}
where
\begin{equation}
\label{eq:A_vec_nu_H1}
    \begin{pmatrix}
        \nu_1 \\ \nu_2 \\ \nu_3 \\ \nu_4 \\ \nu_5 \\ \nu_6
    \end{pmatrix}
    \equiv
    \begin{pmatrix}
        \sev{\frac{1}{2}(\hat{x} - bD_x)^2} \\ \sev{\frac{1}{2} \hat{p}^2} \\ \sev{\hat{p}(\hat{x}-bD_x) + (\hat{x}-bD_x)\hat{p}} \\ \sev{(\hat{x}-bD_x)D_x} \\ \sev{\hat{p}D_x} \\ \sev{D_x^2}
    \end{pmatrix}.
\end{equation}
Note that, for $b = 1$, the quantities $(\nu_1, \nu_2, \nu_3)$ evolve under a closed system of equations. Later, this is used to calculate the asymptotic energy when the system is connected to a thermal bath, see Eq.~\eqref{eq:A_thermal_bath_H1}.

The matrix appearing in Eq.~\eqref{eq:A_H2_matrix_form} has six eigenvalues, but no closed-form solution for arbitrary values of the parameters was obtained. However, in the limit $b \rightarrow 1$, the six eigenvalues can be obtained, to first order in $(1-b)$:
\begin{equation}
    \left( -2\gamma(1-b), -\gamma \pm \sqrt{\gamma^2 - 4\omega^2} - \frac{4\gamma\omega^2(1-b)}{\gamma^2-4\omega^2 \mp \gamma\sqrt{\gamma^2-4\omega^2}}, -b\gamma, -\frac{\gamma}{2} \pm \sqrt{\frac{\gamma^2}{4} - \omega^2} + \frac{2\gamma\omega^2(1-b)}{\gamma^2-4\omega^2 \pm \sqrt{\gamma^2 - 4\omega^2}} \right).
\end{equation}
We also have the approximate solution for these six eigenvalues in the limit $b \rightarrow 0$:
\begin{equation}
    \left( -2\gamma + \frac{2\gamma\omega^2b}{\gamma^2 + \omega^2}, \pm 2i \omega \mp \frac{i\gamma\omega b}{\gamma \pm i\omega}, -\frac{\gamma\omega^2 b}{\gamma^2 + \omega^2}, -\gamma \pm i\omega \mp \frac{i\gamma\omega b}{2(\gamma + \mp i\omega)} \right).
\end{equation}
Note that all eigenvalues have negative real parts for $b \lesssim 1$ and $b \gtrsim 0$; thus, the system converges. We assume the system will converge for all values in the range $0 < b \leq 1$, similar to Protocol XP.

Even though we don't have a closed-form solution for the eigenvalues of the matrix in the most general case, we can still find the analytical results for the fixed point of the dynamics,
\begin{equation}
\label{eq:A_asymptotic_sol_protX}
    \vec{\nu}_\infty = \left( \frac{b\gamma}{16\lambda} - \frac{\gamma\lambda}{4\omega^2} + \frac{\lambda(\gamma^2+\omega^2)}{4b\gamma\omega^2}, \frac{b\gamma}{16\lambda} + \frac{\lambda(\gamma^2+\omega^2)}{4b\gamma\omega^2}, \frac{\lambda}{2\omega}, \frac{\gamma\lambda}{2b\omega^2}, -\frac{\lambda}{2b\omega}, \frac{4\gamma\lambda^2+b\gamma\omega^2}{8b(1-b)\lambda\omega^2} \right)^T.
\end{equation}
The asymptotic energy and position variance results can be calculated directly from Eq.~\eqref{eq:A_asymptotic_sol_protX},
\begin{subequations}
    \begin{align}
        &\sev{\hat{H}_x}_\infty = \hbar\omega\left( \nu_{1,\infty} + \nu_{2,\infty} \right), \\
        &\sev{\hat{x}^2}_\infty = 2\nu_{1,\infty} + 2b\nu_{2,\infty} + b^2\nu_{4,\infty}.
    \end{align}
\end{subequations}
Here, $\nu_{i,\infty}$ for $i=1,\dotsc,6$, denotes the fixed point of the dynamics. Note that $\sev{\hat{x}^2}_\infty$ is equal to the variance of position  since these dynamics give $\sev{\hat{x}}_\infty = 0$ (for $b < 1$).

\section{Protocol XP: Position and Momentum Measurements}\label{sec:A-2measurements}
To solve for the asymptotic expectation value of $\hat{H}_{xp}({\bf D})$, we first use Eq.~\eqref{eq:QFPME_xp} to write a set of evolution equations for $\sev{\hat{H}_{xp}({\bf D})}$ and quantities that couple to it. In matrix form, these equations are:
\begin{equation}
    \label{eq:A_H2_matrix_form}
    \partial_t\begin{pmatrix}
        \nu_1 \\ \nu_2 \\ \nu_3 \\ \nu_4
    \end{pmatrix}
    = \begin{pmatrix}
        -2\gamma b & -\gamma b(b-1) & 0 & 0 \\
        2\gamma & -\gamma & \omega & -\gamma b(b-1) \\
        0 & -\omega & -\gamma & 0 \\
        0 & 2\gamma & 0 & 2\gamma (b-1)
    \end{pmatrix}
    \begin{pmatrix}
        \nu_1 \\ \nu_2 \\ \nu_3 \\ \nu_4
    \end{pmatrix}
    + \begin{pmatrix}
        b^2\frac{\gamma^2}{4\lambda} \\ -b\frac{\gamma^2}{2\lambda} \\ 0 \\ \frac{\gamma^2}{2\lambda}
    \end{pmatrix},
\end{equation}
where
\begin{equation}
    \begin{pmatrix}
        \nu_1 \\ \nu_2 \\ \nu_3 \\ \nu_4
    \end{pmatrix}
    \equiv
    \begin{pmatrix}
        \frac{1}{2}\left[ \sev{(\hat{p}-D_p)^2} + \sev{(\hat{x}-D_x)^2} \right] \\ \sev{(\hat{x}-bD_x)D_x + (\hat{p}-bD_p)D_p} \\ \sev{(\hat{p}-bD_p)Dx - (\hat{x}-bD_x)D_p} \\ \sev{D_x^2 + D_p^2}
    \end{pmatrix}.
\end{equation}
Note that if $b = 1$, the evolution of the average energy, described by $\nu_1$, follows a first-order ODE. Thus, the average energy decays exponentially to its asymptotic value. The relaxation rate in this case is given by $2\gamma$.

The matrix appearing in Eq.~\eqref{eq:A_H2_matrix_form} has four eigenvalues:
\begin{equation}
\label{eq:H2_relaxation_eigenvalues}
    -\gamma \pm \frac{\Delta_{\pm}}{\sqrt{2}},
\end{equation}
where $\Delta_\pm \,=\, \sqrt{-\omega^2+\gamma^2 \,\pm \, \Omega}$ and $\Omega \,=\, \sqrt{\omega^4+2 \left(1-8b+8b^2\right) \omega^2 \gamma ^2+\gamma^4}$. These eigenvalues, which determine relaxation rates, must be negative for the solution of Eq.~\eqref{eq:A_H2_matrix_form} to converge. This condition sets the bounds $0 < b \leq 1$, as expected. 
The asymptotic solution for the system can be found from the fixed point of the dynamics,
\begin{equation}
    \label{eq:A_asymptotic_sol_protXP}
    \vec{\nu}_\infty = \left( \frac{b\gamma}{8\lambda} - \frac{\gamma\lambda}{2\omega^2} + \frac{\lambda(\gamma^2 + \omega^2)}{2b\gamma\omega^2}, \frac{\gamma\lambda}{b\omega^2}, -\frac{\lambda}{b\omega}, \frac{4\gamma\lambda^2 + b\gamma\omega^2}{4b(1-b)\lambda\omega^2} \right)^T.
\end{equation}
The asymptotic energy and position variance results can be calculated directly from Eq.~\eqref{eq:A_asymptotic_sol_protXP},
\begin{subequations}
    \begin{align}
        &\sev{\hat{H}_{xp}}_\infty = \hbar\omega \nu_{1,\infty}, \\
        &\sev{\hat{x}^2}_\infty = \nu_{1,\infty} + b\nu_{2,\infty} + \frac{b^2}{2}\nu_{4,\infty}.
    \end{align}
\end{subequations}
Here, $\nu_{i,\infty}$ for $i=1,\dotsc,4$, denotes the fixed point of the dynamics.

Note that, to derive the equation for $\sev{\hat{x}^2}_\infty$, we explicitly assumed that $\sev{\hat{x}^2}_\infty = \sev{\hat{p}^2}_\infty$ due to the $x \leftrightarrow p$ symmetry in the system. Also, recall that this protocol has a fixed point $\sev{\hat{x}}_\infty = 0$ (for $b < 1$), thus we can write $\sigma_{x,\infty}^2 \equiv \sev{\hat{x}^2}_\infty - \sev{\hat{x}}_\infty^2 = \sev{\hat{x}^2}_\infty$.

\section{Protocol C: Cross Feedback}\label{sec:A-cross_feedback}

To derive the results presented in Sec.~(\ref{sec:cold_damping}) we first define a six-dimensional vector $\vec{\nu}_c=\, (\frac{\omega}{2}\sev{\hat{x}^2}, \,\frac{\omega}{2}\sev{\hat{p}^2},\,\frac{\omega}{2}\sev{(\hat{x}\hat{p}+\, \hat{p}\hat{x} )},\, \sev{D_x \hat{x}}, \,\sev{D_x \hat{p}},\, \sev{D_x^2} )$ to describe the dynamics of average quantities of interest, as follows: 
\begin{equation}
    \partial_t \, \vec{\nu}_c(\omega,\gamma,\lambda,\Gamma,t)=\, \textbf{M}_c(\omega,\gamma,\mu,\Gamma)\,\vec{\nu}_c(\omega,\gamma,\lambda,\Gamma,t)  \,+\, \vec{\nu}_{s,c} \, (\omega,\gamma,\lambda,\Gamma),
    \label{eq:ODE Linear}
\end{equation}
where 
\begin{equation}
    \vec{\nu}_{s,c}(\omega,\,\gamma, \lambda,\Gamma) \,=\, \left(\frac{\Gamma\omega(\bar{n}+1/2)}{2}, \frac{\omega \lambda}{2} + \frac{\Gamma\omega(\bar{n}+1/2)}{2}, 0, 0, 0, \frac{\gamma^2}{4\lambda} \right)^T,
    \label{eq:nu source c}
\end{equation}
and

\begin{align}
\textbf{M}_c(\omega,\,\gamma,\,\mu,\,\Gamma)=\left(
\begin{array}{cccccc}
 -\Gamma & 0 & \omega  & -\mu  \omega ^2 & 0 & 0 \\
 0 & -\Gamma & -\omega  & 0 & \frac{\Gamma\omega\mu}{2} & 0 \\
 -2 \omega  & 2 \omega  & -\Gamma & \frac{\Gamma\omega\mu}{2} & -\mu  \omega ^2 & 0 \\
 \frac{2 \gamma }{\omega } & 0 & 0 & -\gamma -\frac{\Gamma}{2} & \omega  & -\mu  \omega  \\
 0 & 0 & \frac{\gamma }{\omega } & -\omega  & -\gamma -\frac{\Gamma}{2} & \frac{\Gamma\mu}{2} \\
 0 & 0 & 0 & 2 \gamma  & 0 & -2 \gamma  \\
\end{array}
\right).
\label{eq: Ml matrix}
\end{align}

It is generally complicated to solve Eq.\,\eqref{eq:ODE Linear} for an arbitrary set of parameters. However, one can obtain the asymptotic average energy in the case $\Gamma = 0$ by considering $\partial_t \, \vec{\nu}_c(\omega,\gamma,\lambda,t\to \infty)=0$. We then have 
\begin{equation}
    \vec{\nu}_c(\omega,\gamma,\lambda,0,t\to \infty)=\left( \frac{\mu  \omega ^2}{16 \lambda }+\frac{\lambda  (\gamma ^2+\omega ^2)}{4 \gamma ^2 \mu },\frac{\mu  \omega
^2}{16 \lambda }+\frac{\lambda  (\gamma ^2 (1+\mu ^2)-\gamma  \mu  \omega +\omega ^2)}{4 \gamma ^2 \mu },\frac{\lambda }{2},\frac{\lambda
}{2 \mu  \omega },\frac{ \lambda}{2} (-\frac{1}{\gamma  \mu }+\frac{1}{\omega }),\frac{\gamma }{8 \lambda }+\frac{\lambda }{2 \mu  \omega
}\right)^T.
\label{eq:nu steadystate LF}
\end{equation}
Using Eq.\,\eqref{eq:nu steadystate LF}, the asymptotic value for the average internal energy can be calculated using the Hamiltonian in Eq.~\eqref{eq:Hc}:
\begin{equation}
    \label{eq:A_Hc_infty}
    \sev{\hat{H}_c}_\infty = \frac{\hbar\omega}{2}\left[ \frac{\mu\omega}{4\lambda} + \frac{\lambda}{\mu\omega} + \frac{\lambda}{2\gamma} + \frac{\lambda\omega}{\gamma^2\mu} + \frac{\gamma\mu^2}{8\lambda} \right]
\end{equation}

To get the timescale of the process, we need to solve for the eigenvalues of Eq.~\eqref{eq: Ml matrix}. That cannot be done analytically for arbitrary values of $(\omega,\gamma,\lambda,\mu)$, but one can find the eigenvalues when $\mu \rightarrow 0$, which is the condition for ground state cooling. The six eigenvalues, up to $\mathcal{O}(\mu)$, are
\begin{equation}
    \label{eq:eigenvalues_cold_damping}
    \left( -\frac{\gamma^2 \omega \mu}{\gamma^2 + \omega^2}, -\gamma - i\omega + \frac{\gamma \omega \mu}{2(\gamma + i\omega)}, -\gamma + i\omega + \frac{\gamma \omega \mu}{2(\gamma - i\omega)}, -2\gamma + \frac{2\gamma^2 \omega \mu}{\gamma^2 + \omega^2}, -2i \omega - \frac{\gamma \omega \mu }{\gamma - i\omega}, 2i\omega - \frac{\gamma \omega \mu}{\gamma + i\omega} \right).
\end{equation}
If we assume $\gamma \gg \omega \gg \lambda$, then the slowest timescale is $\omega \mu$, which in the optimal condition is $\omega \mu^* = 2\lambda$.

When the bath coupling is non-zero $\Gamma \neq 0$, one can solve the same set of equations to find the asymptotic average energy
\begin{equation}
    \label{eq:A_c_Hss_bath}
    \sev{\hat{H}_{c}}_\infty = \frac{\hbar\left(
    \begin{aligned}
        &2\mu^3\omega^2\gamma^3 + \mu^2 \omega \gamma^2 \left[ 4\omega^2 + \Gamma(2\gamma + \Gamma) \right] + \\
        &8\mu\omega^2\gamma\lambda^2 + 4\omega\lambda^2\left[ 4\omega^2 + (2\gamma + \Gamma)^2 \right] + \\
        &8\Gamma \left[ 4\omega^2 + 4\mu\omega\gamma + \gamma^2(4 + \mu^2) + 4\gamma\Gamma + \Gamma^2 \right] \lambda U_{\text{th}} / \hbar
    \end{aligned}
    \right)}{32\mu\omega\gamma\lambda(\gamma+\Gamma) + 8\lambda\Gamma\left[ 4\omega^2 + (2\gamma+\Gamma)^2 \right]}.
\end{equation}

Let's assume that $\gamma \gg \omega \gg \lambda\gamma/\omega, \mu\gamma, \Gamma\gamma/\omega$. In this case, Eq.\,\eqref{eq:A_c_Hss_bath} becomes
\begin{equation}
    \label{eq:A_c_Hss_bath big gamma omega}
    \sev{\hat{H}_{c}}_\infty = \frac{\mu\omega \left\{ \frac{\hbar\omega}{2} \left[ \frac{\lambda}{\mu\omega} + \frac{\mu\omega}{4\lambda}\right] \right\} + \Gamma U_{\text{th}}} 
    {\mu\omega + \Gamma}.
\end{equation}

\subsection{Position spreading}
To investigate the motion and trapping caused by the cross-feedback protocol, we need to calculate the asymptotic position variance, $\sigma_{x,\infty}^2 \equiv \langle \hat{x}^2 \rangle_\infty - \langle \hat{x} \rangle^2_\infty$. From Eq.~\eqref{eq:nu steadystate LF} we have
\begin{equation}
    \label{eq:A_Hc_x2}
    \langle \hat{x}^2 \rangle_\infty =  \frac{\mu  \omega}{8\lambda }+\frac{\lambda  (\gamma ^2+\omega ^2)}{2 \gamma ^2 \mu \omega }.
\end{equation}
We can determine $\langle \hat{x} \rangle_\infty$ by solving the system of equations
\begin{align}
    &\partial_t \langle \hat{x} \rangle = \omega \left( \langle \hat{p} \rangle - \mu \langle D_x \rangle \right), \\
    &\partial_t \langle \hat{p} \rangle = -\omega \langle \hat{x} \rangle, \\ 
    &\partial_t \langle D_x \rangle = \gamma \left( \langle \hat{x} \rangle - \langle D_x \rangle \right).
\end{align}
The asymptotic solution of this system of equations is:
\begin{equation}
    \langle \hat{x} \rangle_\infty = 0, \quad \langle \hat{p} \rangle_\infty = 0, \quad \langle D_x \rangle_\infty = 0.
\end{equation}

Combining both solutions, we conclude the asymptotic variance of the particle is
\begin{equation}
    \label{eq:A_variance_position_Hc}
    \sigma_{x,\infty}^2 = \frac{\mu  \omega}{8\lambda }+\frac{\lambda  (\gamma ^2+\omega ^2)}{2 \gamma ^2 \mu \omega }.
\end{equation}

 \section{Thermal coupling}\label{sec:A-thermal_coupling}
Here, we present the results for the asymptotic average energy for Protocols X and XP in the simplest case of $b = 1$. Using this simplification, the equations for the asymptotic energy are more symmetrical and easier to solve. We will re-derive the equations as they differ from the previous sections.

When focusing on cooling, the average energy for Protocol X evolves under the following system of linearly coupled differential equations:
\begin{subequations}
\label{eq:A_thermal_bath_H1}
    \begin{align}
        \label{eq:1ode_V}
        &\partial_t \sev{\hat{V}} = \frac{\hbar\omega^2}{2}\langle \hat{G} \rangle -2\gamma  \sev{\hat{V}} + \frac{\hbar\omega \gamma^2}{8\lambda} - \Gamma \langle\hat{V}\rangle + \Gamma\frac{\hbar \omega}{2}(\bar{n}+1/2),\\
        \label{eq:1ode_K}
        &\partial_t \sev{\hat{K}} = -\frac{\hbar\omega^2}{2}\langle \hat{G} \rangle + \frac{\hbar \omega \lambda}{2} - \Gamma \langle\hat{K}\rangle + \Gamma\frac{\hbar\omega}{2}(\bar{n}+1/2),\\
        \label{eq:1ode_xDppxD}
        &\partial_t\langle \hat{G} \rangle = \frac{4}{\hbar} \sev{\hat{K}} - \frac{4}{\hbar} \sev{\hat{V}} - \gamma\langle \hat{G} \rangle - \Gamma\sev{\hat{G}},\\
        \label{eq:1ode_H}
        &\partial_t\langle \hat{H}_x \rangle = \frac{\hbar \omega \lambda}{2} + \frac{\hbar \omega \gamma^2}{8\lambda} - 2\gamma \sev{\hat{V}} - \Gamma \sev{\hat{H}_x} + \Gamma\hbar\omega(\bar{n}+1/2),
    \end{align}
\end{subequations}
where $\hat{V} = \hbar\omega(\hat{x}-D_x)^2/2$, $\hat{K} = \hbar\omega \hat{p}^2/2$ and $\hat{G} = \left( \hat{x} - D_x \right) \hat{p} + \hat{p}\left(  \hat{x} - D_x \right)$. This system can be obtained from Eqs.~\eqref{eq:ev_QD},  \eqref{eq:QFPME_x} and \eqref{eq:bath_a1}.

When $\Gamma \neq 0$, the asymptotic energy relaxes to
\begin{equation}
\label{eq:A_x_Hs_bath}
    \sev{\hat{H}_{x}}_\Gamma = \frac{\sev{\hat{H}_{x}}_\infty\left[ 4\omega^2\gamma + \gamma\Gamma(\gamma+\Gamma) \right] + U_{\text{th}}\left[ 4\omega^2\Gamma +\Gamma(\gamma+\Gamma)^2 \right] + \hbar\omega\lambda\gamma\Gamma \left[ 1 - \frac{\gamma(\gamma+\Gamma)}{4\omega^2} \right]}{4\omega^2(\gamma+\Gamma) + (\gamma+\Gamma)(2\gamma+\Gamma)\Gamma},
\end{equation}
with $U_{\text{th}}$ given by \eqref{eq:thermal_energy}, and $\sev{\hat{H}_{x}}_\infty$ given by Eq.~\eqref{eq:energy-H1} with $b = 1$.

We can proceed similarly for Protocol XP. One can use Eqs.~\eqref{eq:QFPME_xp} and \eqref{eq:bath_a2} to write down a system of linearly coupled ODEs for relevant expectation values. For simplicity, we set $b = 1$, $\lambda_x = \lambda_p = \lambda$ and $\gamma_x = \gamma_p = \gamma$, obtaining
\begin{subequations}
\label{eq:A-2ode}
    \begin{align}
        \label{eq:2ode_V}
        &\partial_t \sev{\hat{V}} = \frac{\hbar\omega^2}{2}\langle \hat{G} \rangle -2\gamma \sev{\hat{V}} + \frac{\hbar \omega \lambda}{2} + \frac{\hbar\omega \gamma^2}{8\lambda} + \Gamma\left( U_{\text{th}}/2 - \sev{\hat{V}} \right),\\
        \label{eq:2ode_K}
        &\partial_t \sev{\hat{K}} = -\frac{\hbar\omega^2}{2}\langle \hat{G} \rangle - 2\gamma \sev{\hat{K}} + \frac{\hbar \omega \lambda}{2} + \frac{\hbar\omega \gamma^2}{8\lambda} + \Gamma\left( U_{\text{th}}/2 - \sev{\hat{K}} \right),\\
        \label{eq:2ode_G}
        &\partial_t \sev{\hat{G}} = \frac{4}{\hbar} \sev{\hat{K}} - \frac{4}{\hbar} \sev{\hat{V}} - 2\gamma\langle \hat{G} \rangle - \Gamma\sev{\hat{G}},\\
        \label{eq:2ode_H2}
        &\partial_t\sev{\hat{H}_{xp}} = (2\gamma+\Gamma)\left[ \frac{2\gamma \sev{\hat{H}_{xp}}_\infty + \Gamma U_{\text{th}}}{2\gamma+\Gamma} - \sev{\hat{H}_{xp}} \right].
    \end{align}
\end{subequations}
Here, $\sev{\hat{H}_{xp}}_{xp}$ is given by Eq.~\eqref{eq:energy-H2} with $b = 1$, $U_{\text{th}}$ is the thermal energy the system would relax to in the absence of measurement and feedback, see Eq.~\eqref{eq:thermal_energy}; $\sev{\hat{K}} = (\hbar\omega/2) \sev{(\hat{p}-D_p)^2}$; $\sev{\hat{V}} = (\hbar\omega/2) \sev{(\hat{x}-D_x)^2}$; and $\sev{\hat{G}} = \sev{(\hat{x}-D_x)(\hat{p}-D_p) + (\hat{p}-D_p)(\hat{x}-D_x)}$. From Eq.~\eqref{eq:A-2ode}, we obtain the asymptotic value of the average energy:
\begin{equation}
    \label{eq:A_H2_heat_bath}
    \sev{\hat{H}_{xp}}_\Gamma = \frac{2\gamma \sev{\hat{H}_{xp}}_\infty + \Gamma U_{\text{th}}}{2\gamma+\Gamma}.
\end{equation}

\section{Trajectory simulation}\label{sec:A_trajectory}
Suppose one starts from the Belavkin equation in Eq.~\eqref{eq:belavkin} and writes the time evolution of the expectation value of powers of position or momentum ($\sev{\hat{x}}, \sev{\hat{x}^2}, \sev{\hat{p}}, \sev{\hat{p}^2}$, etc). These equations form a non-closing system because the evolution of the $n-$th power depends on the $(n+1)-$th power. However, using the Gaussian wave packet assumption \cite{PhysRevA.57.2301,PhysRevA.60.2700,doi:10.1080/00107510601101934} one can derive the following relations:
\begin{subequations}
    \label{eq:gaussian_wave}
    \begin{align}
        &\langle \hat{x}^3 \rangle = \langle \hat{x} \rangle^3 + 3\langle \hat{x} \rangle V_x, \\ 
        &\langle \hat{p}^3 \rangle = \langle \hat{p} \rangle^3 + 3\langle \hat{p} \rangle V_p, \\
        &\langle \hat{x}\hat{p}^2 + \hat{p}^2 \hat{x} \rangle = 2\langle \hat{x} \rangle \langle \hat{p}^2 \rangle + 2\langle p \rangle C, \\ 
        &\langle \hat{x}^2\hat{p} + \hat{p} \hat{x}^2 \rangle = 2\langle \hat{x}^2 \rangle \langle \hat{p} \rangle + 2\langle \hat{x} \rangle C.
    \end{align}
\end{subequations}
Here $V_x = \langle \hat{x}^2 \rangle - \langle \hat{x} \rangle^2$, $V_p = \langle \hat{p}^2 \rangle - \langle \hat{p} \rangle^2$ and $C = \langle \hat{x}\hat{p} + \hat{p}\hat{x} \rangle - 2\langle \hat{x} \rangle \langle \hat{p} \rangle$. Using this assumption, one can close the system of equations and arrive at the following coupled ordinary stochastic differential system of equations:
\begin{subequations}
    \label{eq:SDE-belavkin-general}
    \begin{align}
        &\begin{aligned}
            d\langle \hat{x} \rangle = \omega\left( \langle \hat{p} \rangle - b_xD_x - b_pD_p \right)dt + 2\sqrt{\lambda_x}V_x dW_x + \sqrt{\lambda_p} C dW_p,
        \end{aligned}\\ 
        &\begin{aligned}
            d\langle \hat{p} \rangle = -\omega\left( \langle \hat{x} \rangle - c_xD_x - c_p D_p \right)dt + \sqrt{\lambda_x}C dW_x + 2\sqrt{\lambda_p} V_p dW_p,
        \end{aligned}\\
        \label{eq:SDE-Vx}
        &dV_x = (\omega C + \lambda_p -4\lambda_x V_x^2 - \lambda_p C^2)dt, \\
        \label{eq:SDE-Vp}
        &dV_p = (-\omega C + \lambda_x -4\lambda_p V_p^2 - \lambda_x C^2)dt, \\
        \label{eq:SDE-C}
        &dC = \left[ 2\omega V_p - 2\omega V_x -4\left( \lambda_x V_x + \lambda_p V_p \right)C \right]dt, \\
        &dD_x = \gamma_x \left( \langle \hat{x} \rangle - D_x \right)dt + \frac{\gamma_x}{\sqrt{4\lambda_x}} dW_x, \\ 
        &dD_p = \gamma_p \left( \langle \hat{p} \rangle - D_p \right)dt + \frac{\gamma_p}{\sqrt{4\lambda_x}} dW_p.
    \end{align}
\end{subequations}
The equation for $dD_x$ and $dD_p$ follows from the exponential filtering described in Eq.~\eqref{eq:A_evolution_dD}. This system can be solved numerically using standard algorithms for solving stochastic differential equations. From this solution we can calculate the energy at all times along the trajectory using Eq.~\eqref{eq:general_H}:
\begin{equation}
    \langle \hat{H} \rangle = \frac{\hbar \omega}{2} \left[ V_x + \left( \langle \hat{x} \rangle - c_x D_x - c_p D_p \right)^2 + V_p + \left( \langle \hat{p} \rangle - b_x D_x - b_p D_p \right)^2 \right].
\end{equation}

Note that Eqs.~(\ref{eq:SDE-Vx}, \ref{eq:SDE-Vp}, \ref{eq:SDE-C}) are deterministic, and if $\lambda_x = \lambda_p = \lambda$ they reach symmetric asymptotic values
\begin{equation}
\label{eq:symmetridAsymptoticGaussian}
    (V_x)_\infty = \frac{1}{2}, \quad (V_p)_\infty = \frac{1}{2}, \quad C_\infty = 0.
\end{equation}
\end{document}